\documentclass[conf,hidelinks]{new-aiaa}
\usepackage[utf8]{inputenc}

\usepackage{graphicx}

\usepackage{amsmath}
\usepackage[version=4]{mhchem}
\usepackage{siunitx}
\usepackage{comment}
\usepackage{multirow}
\usepackage{tabularx}
\usepackage{booktabs}
    \newcolumntype{L}{>{\raggedright\arraybackslash}X}
\usepackage{array, makecell}
\usepackage{longtable,tabularx}
\title{Incorporating basic RANS calibrations in existing machine-learned turbulence modeling}

\author{Jiaqi J.L. Li\footnote{PhD Candidate, Mechanical Engineering}}
\affil{Pennsylvania State University, State College, Pennsylvania, USA, 16802}
\author{Yuanwei Bin\footnote{PhD Candidate, Mechanical Engineering}}
\affil{Pennsylvania State University, State College, Pennsylvania, USA, 16802}
\affil{College of Engineering, Peking University, Beijing, China, 100871}
\affil{State Key Laboratory for Turbulence and Complex Systems, Peking University, Beijing, China, 100871}
\author{George P. Huang\footnote{Professor, Mechanical Engineering}}
\affil{Wright State University, Dayton, Ohio, USA, 45435}
\author{Xiang I.A. Yang\footnote{Assistant Professor, Mechanical Engineering}}
\affil{Pennsylvania State University, State College, Pennsylvania, USA, 16802}

\begin{document}

\maketitle

\begin{abstract}
This work aims to incorporate basic calibrations of Reynolds-averaged Navier-Stokes (RANS) models as part of machine learning (ML) frameworks. 
The ML frameworks considered are tensor-basis neural network (TBNN), physics-informed machine learning (PIML), and field inversion \& machine learning (FIML) in {\it J. Fluid Mech.}, 2016, 807, 155-166, {\it Phys. Rev. Fluids}, 2017, 2(3), 034603 and {\it J. Comp. Phys.}, 2016, 305, 758-774, and the baseline RANS models are the one-equation Spalart-Allmaras model, the two-equation $k$-$\omega$ model, and the seven-equation Reynolds stress transport models. 
ML frameworks are trained against plane channel flow and shear-layer flow data.
We compare the ML frameworks and study whether the machine-learned augmentations are detrimental outside the training set. 
The findings are summarized as follows. The augmentations due to TBNN are detrimental. PIML leads to augmentations that are beneficial inside the training dataset but detrimental outside it. These results are not affected by the baseline RANS model. FIML’s augmentations to the two eddy viscosity models, where an inner-layer treatment already exists, are largely neutral. Its augmentation to the seven-equation model, where an inner-layer treatment does not exist, improves the mean flow prediction in a channel. Furthermore, these FIML augmentations are mostly non-detrimental outside the training dataset. 
In addition to reporting these results, the paper offers physical explanations of the results. 
Last, we note that the conclusions drawn here are confined to the ML frameworks and the flows considered in this study. More detailed comparative studies and validation \& verification studies are needed to account for developments in recent years. 
\end{abstract}

\section{Nomenclature}

\renewcommand\arraystretch{1.0}
\noindent\begin{longtable*}{@{}l @{\quad=\quad} l@{}}
$b_{ij}$  & anisotropy tensor part of the normalized Reynolds stress tensor \\
$C_{b1}$, $C_{b2}$, $C_{w1}$, $C_{w2}$, $C_{w3}$ & Spalart-Allmaras model constants\\
$d$ & distance to the wall\\
$D_{ij}$ & diffusion tensor in the Reynolds stress transport equations\\
${\rm D}/{\rm D}t$ & material derivative\\
$f_{\nu1}$, $f_{\nu2}$, $f_w$, $g$ & Spalart-Allmaras model functions\\
$I_2$, $I_3$ & identity matrices\\
$k$ & turbulent kinetic energy\\
$\mathcal{P}_{ij}$ & production tensor in the Reynolds stress transport equations\\
$r$ & Spalart-Allmaras model variable\\
$R_{ij}$  & Reynolds stress tensor \\
$Re_\tau$ & friction Reynolds number\\
$S_{ij}$, $\boldsymbol{S}$ & strain rate tensor\\
$u$, $U$ & velocity\\
$x_i$ & $i$-th Cartesian direction\\
$y$ & vertical direction, wall-normal direction\\
$\beta$   & model correction \\
$\theta$, $\theta^*$ & Wilcox $k$-$\omega$ model constants\\
$\gamma$ & model constant in the Wilcox $k$-$\omega$ model\\
$\delta$ & channel half height\\
$\delta_{ij}$ & Kronecker delta tensor\\
$\epsilon_{ij}$ & dissipation tensor in the Reynolds stress transport equations\\
$\epsilon$ & dissipation rate\\
$\kappa$ & von K{\'a}rm{\'a}n constant\\
$\mu$ & molecular dynamic viscosity\\
$\mu_t$ & turbulent eddy viscosity\\
$\nu$ & molecular kinematic viscosity\\
$\nu_t$, $\tilde{\nu}$ & turbulent eddy viscosity\\
$\Phi_{ij}$ & pressure-strain correlation tensor in the Reynolds stress transport equations\\
$\sigma$, $\sigma^*$ & Spalart-Allmaras model constants\\
$\tau_{ij}$ & Reynolds stress tensor\\
$\chi$ & Spalart-Allmaraos model variable\\
$\omega$ & specific dissipation rate\\
$\Omega_{ij}$, $\boldsymbol{R}$ & rotation tensor\\

\end{longtable*}

\section{Introduction}
Due to the high computational cost associated with scale-resolving tools \cite{choi2012grid,yang2021grid,li2022grid}, as well as the limited computing power, cost-effective, non-scale-resolving tools like Reynolds averaged Navier Stokes (RANS) will remain the workhorse for fluids engineering for the foreseeable future.
RANS solves for the mean flow while modeling the entirety of turbulence.
Turbulence modeling in the context of RANS is a topic with a long history \cite{pope2000turbulent,durbin2018some,kalitzin2005near}.
Early models are algebraic \cite{smith1967numerical,baldwin1978thin} and have limited applicable ranges.
The work in the 80s and 90s led to a series of transport RANS models that are widely used today.
Notable examples include the one-equation Spalart-Allmaras model \cite{spalart1992one}, the various two-equation $k$-$\omega$ model \cite{wilcox1988reassessment, menter1994two}, the seven-equation full Reynolds stress model \cite{speziale1991modelling,launder1975progress}, among others \cite{chien1982predictions,hanjalic2004robust,durbin1995separated}.
These RANS models, or variations thereof, are readily available in many commercial codes and are the industrial standards.
Although off-the-shelf RANS models are known to have difficulty handling separation, boundary-layer three-dimensionality, and unsteadiness, among others \cite{slotnick2014cfd,slotnick2011overview,rumsey2023fourth},
they generally produce results that are considered ``reasonable'' in the sense that they are usually not apparently unphysical.

The story is different regarding machine-learning models.
The use of machine-learning tools for turbulence modeling dates back to as early as 2002 \cite{milano2002neural}, but their extensive use was not until the work in Refs. \cite{duraisamy2015new,ling2016reynolds} in 2015 and 2016.
Since then, numerous researchers have delved into data-based approaches for turbulence modeling, resulting in promising models that have demonstrated exceptional accuracy for flows within their training sets \cite{parish2016paradigm,singh2017machine,han2022vcnn}. 
However, akin to other machine-learning applications, these models face limitations in their generalization capabilities.
Worse still, when applied to flows outside the training set, machine-learned augmentations are often detrimental \cite{rumsey2022search,spalart2022old}, leading to unphysical results.
This lack of robustness prevents the use of machine-learning models in an engineering setting, which, in turn, undermines the potential impact of endeavors aimed at enhancing learning efficiency \cite{singh2017machine,holland2019towards} and capturing the elusive flow physics absent in conventional models \cite{han2022vcnn,shirian2022eddy,mani2021macroscopic}.

Understanding why machine-learning models do not generalize is crucial. Exploring the answer to the question can help identify the underlying issues and guide the development of machine-learning models.
While a definitive answer to this question is challenging, insights can be gained from the discussions in Refs. \cite{rumsey2022search,spalart2022old,menter2019best,bin2022progressive,bin2023data,vadrot2023survey}.
The basic calibrations of empirical RANS models, such as decaying isotropic homogeneous turbulence, zero-pressure-gradient flat-plate boundary layers, and free shear flows, are the fundamental building blocks that make up a turbulent flow.
These basic calibrations enable empirical RANS models to generalize beyond their calibration datasets.
Based on this observation, one can surmise that machine-learning models struggle to generalize precisely because they do not preserve these basic calibrations.
Incorporating the basic calibrations of RANS models into machine-learning approaches could potentially grant them the generalization abilities observed in empirical RANS models.
An important implication of this explanation is that solely focusing on capturing missing physics in conventional empirical models, as attempted in Refs. \cite{ling2016reynolds,wu2018physics}, will not automatically grant machine-learning models the generalizability of empirical models. 

Attempting to incorporate all basic calibrations of empirical RANS models into machine-learning models is excessively ambitious and unnecessary. 
Among the basic calibrations of conventional empirical models, the law of the wall (LoW), i.e., the scaling of the mean flow as a function of the distance from the wall-normal coordinate in the viscous and logarithmic layer, is regarded by many as the most important \cite{menter2019best, spalart2015philosophies}.
Spalart argued that ``the law of the wall is the most useful and trusted part of our knowledge of turbulence in the field of turbulence modeling'' \cite{spalart2015philosophies}.
Menter and company further emphasized that: while certain aspects such as decaying homogeneous isotropic turbulence and free-shear flows may be disregarded, preserving the law of the wall is imperative \cite{menter2019best}.
Therefore, in order to enable generalization, it is essential to incorporate the law of the wall as an integral component of machine-learning models. This work focuses precisely on the incorporation of the law of the wall into machine-learning models.

We focus on three well-established modeling frameworks: tensor basis neural networks (TBNN)  introduced in Ref. \cite{ling2016reynolds}, physics-informed machine learning (PIML) proposed in Ref. \cite{wang2017physics}, and field inversion and machine learning first formulated in Ref. \cite{parish2016paradigm}.
Below, we provide a summary of these three frameworks, with a more comprehensive discussion deferred to Section \ref{sec:ML_method}.
The objective of TBNN is to model the anisotropic part of the Reynolds stress, denoted as $b_{ij}$.
Ling et al. \cite{ling2016reynolds} invoked the closed-form expression for $b_{ij}$ in Ref. \cite{pope1975more}, which expresses $b_{ij}$ as a function of the strain rate tensor $S_{ij}$ and the rotational rate tensor $\Omega_{ij}$.
A deep neural network is employed to learn the coefficients in the expression as a function of the invariants of $S_{ij}$ and $\Omega_{ij}$.
Consequently, TBNN exhibits Galilean invariance.
PIML, on the other hand, focuses on modeling the discrepancy between the Reynolds stress predicted by RANS and that observed in DNS. This is achieved through a random forest regressor, which takes as inputs all locally available flow information. 
Notably, PIML also maintains Galilean invariance when evolving invariant inputs only.
The objective of FIML is to augment the auxiliary equation in RANS models such that the resulting model yields more accurate predictions of some given quantity of interest.
The method consists of field inversion followed by machine learning. 
During the field inversion step, an augmentation is learned by minimizing a cost function through model-consistent training.
During the machine learning step, a neural network is trained to relate the augmentation obtained from the field inversion step to the flow variables.

TBNN, PIML, and FIML have had further developments since the early publications \cite{ling2016reynolds,wang2017physics,parish2016paradigm}.
Here, we review some of these developments.
Parish et al. \cite{parish2023turbulence} argued that instead of modeling the anisotropic part of the Reynolds stress, modeling the error in the Reynolds stress anisotropy is a more reasonable choice for TBNN. 
Xu et al. and Xie et al. \cite{xu2023artificial,xie2020artificial} extended the framework of TBNN to LES modeling.
Wu et al. \cite{wu2018physics} explored velocity propagation for PIML.
Zhang et al. \cite{zhang2019recent} utilized the ensemble Kalman method to handle sparse data for PIML.
Singh et al. \cite{singh2017machine} incorporated adjoint-based optimization into FIML, which significantly improved the training efficiency.

In addition to the aforementioned frameworks, other machine-learning frameworks for RANS modeling have also emerged.
A notable example is the symbolic regression method in Refs. \cite{weatheritt2016novel,zhao2020rans,fang2023toward}.
There, an evolutionary algorithm drives algebraic forms of the Reynolds stress anisotropic tensor. 
The method is appealing because it gives mathematically interpretable forms.
Yet another framework is progressive machine learning \cite{bin2022progressive,bin2023progressive}.
The method aims to preserve the basic calibrations of a baseline model.
Applying this framework, Bin et al. obtained a data-enabled recalibration of the SA model, maintaining the basic calibrations (e.g., flat plate boundary layer and free shear flows) while improving the model's predictions of flow separation.
Considering the multitude of existing methods, implementing all of them would be extremely time-consuming and may not provide substantial additional insights at this stage. 
Therefore, a comparative study of methods other than TBNN, PIML, and FIML is left for future investigation.

Besides the ML frameworks, we must also pick baseline models.
This study focuses on the one-equation Spalart-Allmaras (SA) model \cite{spalart1992one}, the two-equation Wilcox $k$-$\omega$ model \cite{wilcox1988reassessment}, and the SSG seven-equation full Reynolds stress model (FRSM) \cite{speziale1991modelling}.
Both the SA model and the Wilcox $k$-$\omega$ model are extensively used in the industry and are available in many software codes \cite{fluent2011fluent,weller1998tensorial}.
The Speziale-Sarkar-Gatski (SSG)-FRSM, while less commonly used, is implemented in STARCCM+, OpenFOAM, and ANSYS Fluent.
We postpone further details of these models to section \ref{sec:RANS}.

The remainder of this paper is organized as follows.
Details of the baseline RANS models and the machine-learning frameworks are presented in Sec. \ref{sec:method}.
The results are shown in Sec. \ref{sec:result}.
Last, we conclude in section \ref{sec:conclusion}

\section{Methodology}
\label{sec:method}

We present the details of the baseline RANS models in section \ref{sec:RANS} and the details of the machine-learning frameworks in section \ref{sec:ML_method}.
The details of the training and testing data are presented in section \ref{sec:data}.

\subsection{Baseline RANS models}
\label{sec:RANS}




\subsubsection{Spalart-Allmaras model}

The Spalart-Allmaras (SA) model in Ref. \cite{spalart1992one} is a one-equation model.
It solves a transport equation for $\tilde{\nu} = \nu_t/f_{\nu1}$, where $\nu_t$ is the eddy viscosity, and $f_{\nu1}$ is a function of the dimensionless parameter $\chi \equiv \tilde{\nu}/\nu$.
The transport equation for $\tilde{\nu}$ is given by:
\begin{equation}
\frac{\rm D \tilde{\nu}}{{\rm D} t}=c_{b1} \tilde{S}\tilde{\nu} -c_{w1}f_w\left(\frac{\tilde{\nu}}{d}\right)^2+\frac{1}{\sigma}\left[\frac{\partial }{\partial x_j}\left((\nu+\tilde{\nu})\frac{\partial \tilde{\nu}}{\partial x_j}\right)+c_{b2}\frac{\partial \tilde{\nu}}{\partial x_j}\frac{\partial \tilde{\nu}}{\partial x_j}\right],
\label{eq:SA}
\end{equation}
where ${\rm D}/{\rm D} t$ is the material derivative, $\tilde{S}=\Omega+\tilde{\nu}/(\kappa^2d^2)f_{\nu2}$, $\kappa = 0.41$ is the von K\'arm\'an constant, $\Omega$ is the vorticity magnitude, $c_{w1} = c_{b1}/\kappa^2 + (1+c_{b2})/\sigma$, and $d$ is the distance to the closest wall.
The functions $f_{\nu1}$, $f_{\nu2}$ are 
\begin{equation}
    f_{\nu1} = \frac{\chi^3}{\chi^3+c_{\nu1}^3}, ~~f_{\nu2} = 1 - \frac{\chi}{1+\chi f_{\nu1}};
    \label{eq:fnu12}
\end{equation}
The function $f_w$, which depends on $r = \tilde{\nu}/(\tilde{S}\kappa^2 d^2)$, is given by:
\begin{equation}
    f_w(r) = g\left(\frac{1+c_{w3}^6}{g^6 + c_{w3}^6} \right)^{1/6}, ~~g=r+c_{w2}(r^6-r);
\end{equation}
The six model constants are: $\sigma = 2/3$, $c_{b1} = 0.1355$, $c_{b2} = 0,622$, $c_{w2} = 0.3$, $c_{w3} = 2$, and $c_{\nu1} = 7.1$.

\subsubsection{Wilcox \texorpdfstring{$k$-$\omega$}{TEXT} model}

The Wilcox $k$-$\omega$ model in \cite{wilcox1988reassessment} is a two-equation model.
It solves two equations for $k$ and $\omega$:
\begin{equation}
\begin{split}
\frac{{\rm D} k}{{\rm D} t}&=\tau_{ij}\frac{\partial u_i}{\partial x_j}-\theta^* \omega k +\frac{\partial }{\partial x_j}\left[\left(\nu+\sigma^*\nu_t\right)\frac{\partial k}{\partial x_j}\right],\\
\frac{\rm D \omega}{{\rm D} t}&=\frac{\gamma \omega}{k}\tau_{ij}\frac{\partial u_i}{\partial x_j}-\theta \omega^2 +\frac{\partial }{\partial x_j}\left[\left(\nu+\sigma\nu_t\right)\frac{\partial \omega}{\partial x_j}\right],
\end{split}
\label{eq:kw}
\end{equation}
with
\begin{equation}
\nu_t = k/\omega,
\end{equation}
and
\begin{equation}
R_{ij}=2\nu_t\left(S_{ij}-\frac{1}{3}\frac{\partial u_k}{\partial x_k}\right) - \frac{2}{3} k \delta_{ij}, ~~S_{ij} = \frac{1}{2}\left( \frac{\partial u_i}{\partial x_j} + \frac{\partial u_j}{\partial x_i} \right),
\end{equation}
where $\delta_{ij}$ is the Kronecker delta.
The five model constants are: $\gamma = 5/9$, $\theta^* = 0.09$, $\theta = 0.075$, $\sigma^* = 0.5$, and $\sigma = 0.5$.

\subsubsection{Full Reynolds Stress Model}

FRSM solves the transport equations of the Reynolds stresses
\begin{equation}
\centering
\frac{ {\rm D} R_{ij}}{{\rm D} t}=\mathcal{P}_{ij}-\epsilon_{ij}+\mathcal{D}_{ij}+\Pi_{ij},
\label{eq:FRSM}
\end{equation}
where $\mathcal{P}_{ij}$ is the production tensor defined as $\mathcal{P}_{ij}=-R_{ik}\partial{\overline{u}_j}/\partial{x_k}-R_{jk}\partial{\overline{u}_i}/\partial{x_k}$,  $\epsilon_{ij}$ is the dissipation tensor, $\mathcal{D}_{ij}$ is the diffusion tensor, and $\Pi_{ij}$ is the pressure-strain correlation.
The last three terms on the right-hand side are unclosed.
The dissipation tensor is modeled as \cite{launder1975progress}
\begin{equation}
\epsilon_{ij}=\frac{2}{3}\epsilon\delta_{ij}.
\label{eq:dissipation}
\end{equation}
The dissipation rate, $\epsilon$, is obtained by solving the following transport equation
\begin{equation}
\frac{{\rm D}\epsilon}{{\rm D} t} = C_{\epsilon 1}\frac{\mathcal{P}\epsilon}{k}-C_{\epsilon 2}\frac{\epsilon^2}{k}+\frac{\partial}{\partial x_i} \left[ \left( \nu \delta_{ij} +C_\epsilon\frac{k}{\epsilon} \langle u_iu_j \rangle \right) \frac{\partial \epsilon}{\partial x_j}  \right],
\end{equation}
where $\mathcal{P}$ is the production of TKE defined as $\langle u_iu_j\rangle \partial u_i/\partial x_j$; $C_{\epsilon 1}$, $C_{\epsilon 2}$ and $C_{\epsilon}$ are model constants.
The diffusion term and pressure-strain correlation term are modeled via the Daly-Harlow (DH) gradient diffusion model \cite{daly1970transport} and the SSG model \cite{speziale1991modelling}:
\begin{equation}
    \mathcal{D}_{ij}=\frac{\partial}{\partial x_k}\left(\nu\delta_{kl}+C_{DH}\frac{kR_{kl}}{\epsilon}\right)\frac{\partial R_{ij}}{\partial x_l}
\label{eq:DH}
\end{equation}
and
\begin{equation}
\begin{aligned}
    \Pi_{ij} &= - C_1 \epsilon+C_1^*\mathcal{P}_{kk} b_{ij}+C_2\epsilon\left( b_{ik}b_{kj} - \frac{1}{3}b_{mn}b_{mn}\delta_{ij}\right) + \left(C_3-C_3^*\sqrt{b_{kl}b_{kl}}\right )kS_{ij} \\
    &+C_4k \left( b_{ik}S_{jk}+b_{jk}S_{ik}-\frac{2}{3}b_{mn}S_{mn}\delta_{ij} \right ) + C_5k\left (b_{ik}W_{jk}+b_{jk}W_{ik}\right )
\end{aligned}
\label{eq:SSG}
\end{equation}
The model constants are:
\begin{equation}
\centering
\begin{split}
    C_{\epsilon 1} = 1.44, C_{\epsilon 2} = 1.83, & C_{\epsilon} = 0.15, C_{DH}=0.22\\ 
    C_1=3.4, C_1^*=1.8, C_2=4.2, C_3 =\frac{4}{5}, & C_3^*=1.3, C_4 = 1.25, C_5 = 0.4.
\end{split}
\end{equation}

\vspace{5mm}

Table \ref{tab:RANS} summarizes the relevant information of the baseline models.
Whether the model is an eddy viscosity model and the number of auxiliary transport equations are defining features of an empirical RANS model.
Whether the model contains the turbulent kinetic energy and dissipation rate information determines whether TBNN and PIML are applicable to the baseline model and therefore are relevant to this study.
A wall treatment ensures that the model captures the behavior of the mean flow in the viscous sublayer and the buffer layer.
For the three models considered here, a wall treatment exists in both the SA model and the Wilcox $k$-$\omega$ model;
there is a wall treatment for FRSM in the literature \cite{launder1975progress}, but this is not implemented in the present code.

\begin{table}
\caption{\label{tab:RANS} Details of the RANS models. Here, Y and N stand for Yes and No.}
\centering
\begin{tabular}{lccccc}
\hline
& Eddy viscosity & Equations & Contains k info & Constains $\epsilon$ info & Wall treatment \\ \hline
SA               & Y       & 1     & N    & N    & Y \\
Wilcox $k$-$\Omega$ & Y    & 2     & Y    & Y    & Y \\
SSG FRSM         & N       & 7     & Y    & Y    & N \\ \hline
\end{tabular}
\end{table}

We implement the SA and Wilcox $k$-$\omega$ models in Python.
A second-order central difference is employed in space, and an implicit Euler method is used for time integration.
We validate these two implementations against the ones in OpenFOAM regarding plane channel flow (not shown for brevity).
An implementation of the SSG FRSM model exists in OpenFOAM, and we make use of that implementation.

\subsection{Machine-learning frameworks}
\label{sec:ML_method}

\subsubsection{TBNN}

\begin{figure}
\centering
\includegraphics[width=.8\textwidth]{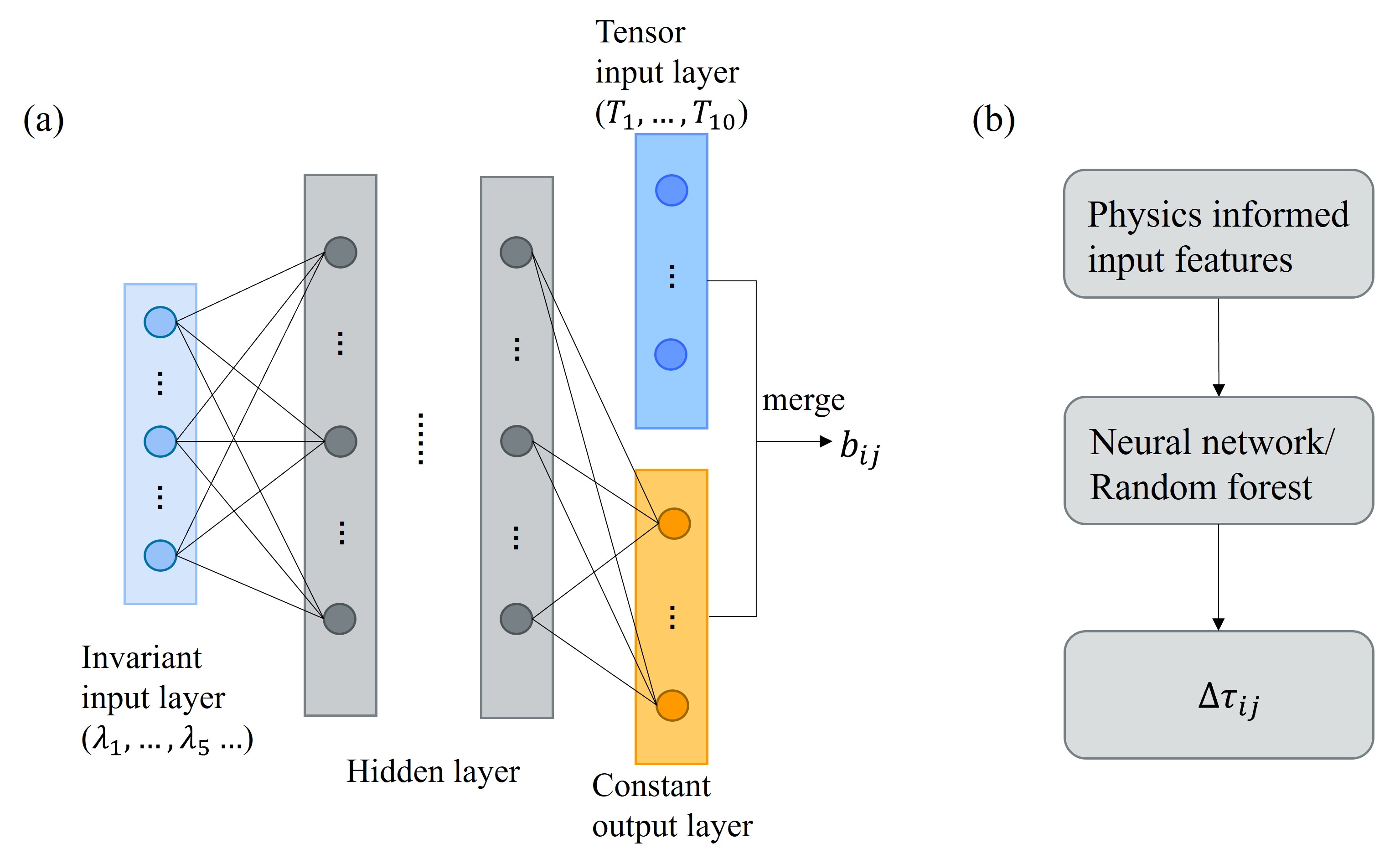}
\caption{Schematic of (a) TBNN and (b) PIML.}
\label{fig:tbnn_piml}
\end{figure}

TBNN, as introduced in Ref. \cite{ling2016reynolds}, predicts the Reynolds stress anisotropy tensor, denoted as $b_{ij} = R_{ij}/2k - (1/3)\delta_{ij}$, following the formulation outlined in Ref. \cite{pope1975more}:
\begin{equation}
b_{ij}=\sum_{k=0}^{10} c_k(\lambda_1,\lambda_2,...,\lambda_5) T^{(k)},
\end{equation}
where
\begin{equation}
\centering
\begin{split}
    T^{(0)} = \frac{1}{3}\boldsymbol{I_3}  - \frac{1}{2}\boldsymbol{I_2},~T^{(1)} = \boldsymbol{S},~& T^{(2)} = \boldsymbol{SR}-\boldsymbol{RS}, ~T^{(3)} = \boldsymbol{S^2}-\frac{1}{3}\boldsymbol{I} \cdot Tr(\boldsymbol{S^2}), ~T^{(4)} = \boldsymbol{R^2}-\frac{1}{3}\boldsymbol{I} \cdot Tr(\boldsymbol{R^2}), \\
    ~T^{(5)} = \boldsymbol{RS^2} - \boldsymbol{S^2R},
    & ~T^{(6)} = \boldsymbol{R^2S} + \boldsymbol{SR^2} -\frac{2}{3}\boldsymbol{I} \cdot Tr(\boldsymbol{SR^2}),
    ~T^{(7)} = \boldsymbol{RSR^2} - \boldsymbol{R^2SR},\\
    ~T^{(8)} = \boldsymbol{SRS^2} - \boldsymbol{S^2RS}, 
    & ~T^{(9)} = \boldsymbol{R^2S^2+S^2R^2} - \frac{2}{3}\boldsymbol{I} \cdot Tr(\boldsymbol{S^2R^2}),
    ~T^{(10)}= \boldsymbol{RS^2R^2} - \boldsymbol{R^2S^2R},
\end{split}
\label{eq:tensor}
\end{equation}  
and
\begin{equation}
\centering
    \lambda_1 = Tr(\boldsymbol {S^2}), ~\lambda_2 = Tr(\boldsymbol {R^2}), ~\lambda_3 = Tr(\boldsymbol {S^3}), ~\lambda_4 = Tr(\boldsymbol {R^2S}), ~\lambda_5 = Tr(\boldsymbol {R^2S^2}).
\label{eq:invariant}
\end{equation}
Here, $T^{(0)}$ is reversed for two-dimensional flow only, $\boldsymbol{S}$ is the strain rate tensor, $\boldsymbol{R}$ is the rotation tensor, $\boldsymbol{I_3}$ is the identity matrix, and $\boldsymbol{I_2}$ is also the identity matrix but $I_{i3}=0$.
For two-dimensional flow, only $T_0$, $T_1$, $T_2$, $\lambda_1$, and $\lambda_2$ are non-zero.
Figure \ref{fig:tbnn_piml} (a) shows a schematic of TBNN.
The network comprises two input layers: the invariant input layer and the
tensor input layer. 
While the invariant input layer can be expanded to include parameters like Reynolds number and distance from the wall \cite{xu2023artificial,guo2021practical}, such expansion is not explored in this study.
We adhere to the setup proposed in Ref. \cite{ling2016reynolds} when configuring our neural network.
The network is a feedforward neural network. 
It has 4 hidden layers with 30 neurons in each layer.
The activation function is leaky-ReLU.
The cost function is the root-mean-square-error (RMSE) in $b_{ij}$.

\subsubsection{PIML}

Physics-informed inputs are fed to a regression tool to predict the error in $R_{ij}$.
The regression tool can take the form of either a neural network or a random forest. The training process relies on benchmark DNS data, and in the work of Wang et al. \cite{wang2017physics}, there are approximately O(10) inputs. 
However, for one-dimensional flows, such as channels and temporally evolving shear layers, the number of inputs reduces to four.
The four inputs $\hat{q}_i$ and their normalizations $q_i^*$ are tabulated in Table \ref{table:piml}.
Following Ref. \cite{wang2017physics}, the inputs to the neural network are $q_\beta=\hat{q}_\beta/(|\hat{q}_\beta|+|q_{\beta}^*|)$ to prevent mathematical singularity.

Concerning the output, PIML in Ref. \cite{wang2017physics} predicts the magnitude $(\Delta k)$, shape $(\Delta \xi,\Delta \eta)$, and orientation $(\Delta \phi_1,\Delta \phi_2,\Delta \phi_3)$ of the object tensor.
For plane channel flow, the output reduces to $(\Delta k,\Delta b_{12})$.
The network is a feedforward neural network. 
It has 4 hidden layers with 30 neurons in each layer.
The activation function is the sigmoid function.

\begin{table}
\caption{\label{tab:table1} Inputs of PIML.}
\centering
\begin{tabular}{llcc}
\hline
Feature$(q_i)$ & Description & Raw input $\hat{q}_i$ & Normalization factor $q_{i}^*$\\\hline
$q_1$& Turbulent kinetic energy & $k$ & $\frac{1}{2}U_iU_i$\\
$q_2$& Wall-distance based Reynolds number & $\sqrt{k}d/50\nu$ & NA \\
$q_3$& Ratio of turbulent time scale to mean strain time scale & $k/\epsilon$ & $1/||\boldsymbol{S}||$\\
$q_4$& Ratio of total to normal Reynolds stresses & $\lVert \langle u_iu_j \rangle \lVert $ & $k$\\
\hline
\end{tabular}
\label{table:piml}
\end{table}

\subsubsection{FIML}

\begin{figure}
\centering
\includegraphics[width=.8\textwidth]{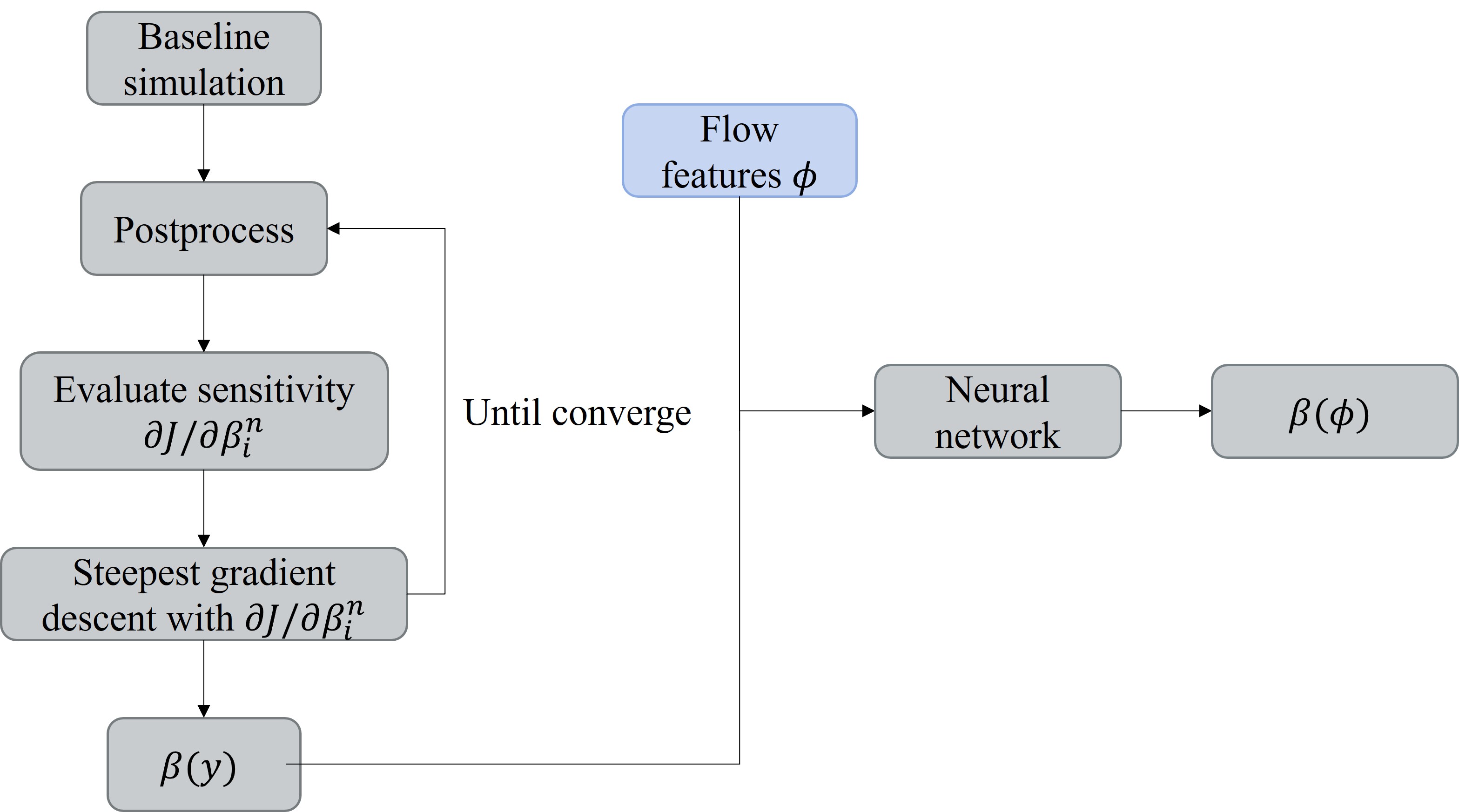}
\caption{Schematic of FIML.}
\label{fig:FIML}
\end{figure}

FIML consists of two steps: field inversion followed by machine learning, as sketched in Fig. \ref{fig:FIML}.
During the field inversion step, an augmentation of the baseline RANS model is learned.
In this study, the augmentation manifests as a multiplier in front of $f_{v1}$ in the SA model, a multiplier in front of the production term in the $k$ equation in the Wilcox $k$-$\omega$ model, and a multiplier in front of the production term in the $\epsilon$ equation in the SSG FRSM, respectively. 
The field inversion involves minimizing the following cost function
\begin{equation}
\centering
J_c(\beta) = \lVert \phi_d - \phi_r(\beta)\lVert_2^2-\lambda\lVert \beta -1.0 \lVert_2^2,
\label{eq:cost}
\end{equation}
where $\phi_d$ is the truth and $\phi_r$ is the RANS prediction.
$\phi$ can be any variable.
For plane channel flow, $\phi=U$ is the mean velocity profile. 
The second term in Eq. \eqref{eq:cost} is a regularization term.
Here, we set $\lambda= 0.001$.
To minimize $J_c$, we resort to gradient descent
\begin{equation}
\centering
\beta^{i+1} = \beta^i - \alpha \frac{\partial J}{\partial \beta_i},
\label{eq:gradient_decsent}
\end{equation}
where gradient $\partial{J}/\partial{\beta_i}$ can be computed via either an adjoint method or a finite difference method.
Here, we employ the adjoint method for the SA and the Wilcox $k$-$\omega$ models and the finite difference method for the SSG FRSM.
Next, a regression tool is used to relate flow features to the augmentation $\beta$.
We use a neural network as our regression tool.
The flow features we use for channel flow are:
\begin{equation}
f_{SA} = \left \{ \frac{\nu}{\nu_t+\nu} \right \},
~f_{k \rm{-} \omega} = \left \{ \frac{\nu}{\nu_t+\nu}, -\frac{\nu}{dp/dx}\frac{dU}{dy} \right \},
~f_{\rm FRSM} = \left \{ \frac{k/\epsilon}{k/\epsilon+\lVert\boldsymbol{S}\lVert}, \frac{\sqrt{k}d}{\nu} \right \},
\label{eq:feature}
\end{equation}
for the SA, the Wilcox $k$-$\omega$, and the SSG FRSM, respectively \cite{rumsey2022search}. 

\vspace{5mm}

Table \ref{tab:ML} summarizes the relevant information of the ML frameworks.
{\it A priori} training is the method where one trains an intermediate variable to match high-fidelity data.
Model consistent training is the method where one trains directly to match the quantity of interest (QoI).
Duraisamy \cite{duraisamy2021perspectives} argued that model-consistent training is preferred over {\it a prior} training because the correct prediction of an intermediate variable does not necessarily guarantee the correct prediction of the end QoI.
``RANS interface'' identifies where a baseline model interfaces with the machine-learning framework.
When the interface is the Reynolds stress, the baseline model is intact, and whether one conducts velocity propagation becomes relevant.
When the interface is the transport equation, the machine-learning augmentation becomes an integral part of the model, and velocity propagation is implied.

\begin{table}
\caption{\label{tab:ML}Details of the ML frameworks. 
}
\centering
\begin{tabular}{lllll}
\hline
     & Target                      & Velocity propagation & RANS Interface     & Training                        \\ \hline
TBNN & Reynolds stress             & N                    & Reynolds stress    &  {\it a priori}  \\
PIML & Error in Reynolds stress    & N                    & Reynolds stress    & {\it a priori}  \\
FIML & Any QoI & N/A                & Transport equation & Model consistent               \\ \hline
\end{tabular}
\end{table}

For all ML frameworks, training is conducted in Python. 
The training involved an Adam optimizer. 
All training does not stop until the cost function cannot be further reduced in another 300 epochs.

\subsection{Data}
\label{sec:data}

We make use of the plane channel flow DNS data in the Johns Hopkins Turbulence Database and the UP Madrid database \cite{moser1999direct,lee2015direct,graham2016web} and the mixing layer DNS data in Ref. \cite{huang2021determining}.

The channel flow data are at the friction Reynolds numbers $\text{Re}_\tau = 180$, $395$, $590$, $1000$, $2000$, and $5200$.
The details of the DNSs are tabulated in Table \ref{tab:POD_data_table}.
The mean flow data should be highly accurate according to Refs. \cite{yang2021grid, oliver2011bayesian}.
For this study, data at $Re_\tau=$180, 590, 1000, and 2000 are used for training; for testing, we use data at $Re_\tau=180$, one of the training flow conditions in the training set, $Re_\tau=395$, a condition that can be obtained by interpolating inside the training set, and $Re_\tau=5200$, a condition that must be obtained through extrapolation. 
Interpolation and extrapolation here are with respect to the flow parameter space, i.e., the Reynolds number, rather than the input space.

\begin{table}
\centering
\caption{Details of the channel flow DNS data. Here, the asterisk and plus superscripts indicate normalization by the channel half-width $\delta$ and the viscous length scale $\nu / u_\tau$, respectively. Further, the subscript $w$ refers to values at the wall.}
\begin{tabular}{ccccc}
\hline
\textbf{\begin{tabular}
[c]{@{}c@{}}Reynolds Number:\\ $ \, $\end{tabular}} & \textbf{\begin{tabular}[c]{@{}c@{}}Domain:\\ $L_x^* , L_y^* , L_z^*$\end{tabular}} & \textbf{\begin{tabular}[c]{@{}c@{}}Grid:\\ $N_x , N_y , N_z$\end{tabular}} & \textbf{\begin{tabular}[c]{@{}c@{}}Resolution:\\ $\Delta x^+ , \Delta y_w^+ , \Delta z^+$\end{tabular}} \\
 \hline
$\text{Re}_\tau = 180$  & $2\pi , 2 , 4\pi/3$   & $128 , 129 , 128$     & $17.7 , 0.05 , 5.9$ \\
$\text{Re}_\tau = 395$  & $4\pi , 2 , \pi$   & $256 , 193 , 192$     & $10.0 , 0.03 , 6.5$ \\
$\text{Re}_\tau = 590$  & $4\pi , 2 , \pi$   & $384 , 257 , 384$     & $9.7 , 0.04 , 7.2$ \\
$\text{Re}_\tau = 1000$ & $8\pi , 2 , 3\pi$  & $2048 , 512 , 1536$    & $10.3 , 0.02 , 4.6$ \\
$\text{Re}_\tau = 2000$ & $8\pi , 2 , 3\pi$  & $4096 , 768 , 3072$    & $8.2 , 0.002 , 4.1$ \\
$\text{Re}_\tau = 5200$ & $8\pi , 2 , 3\pi$  & $10240 , 1536 , 7680$  & $12.7 , 0.07 , 6.4$ \\\hline                                     
\end{tabular}
\label{tab:POD_data_table}
\end{table}

\begin{figure}
\centering
\includegraphics[width=.3\textwidth]{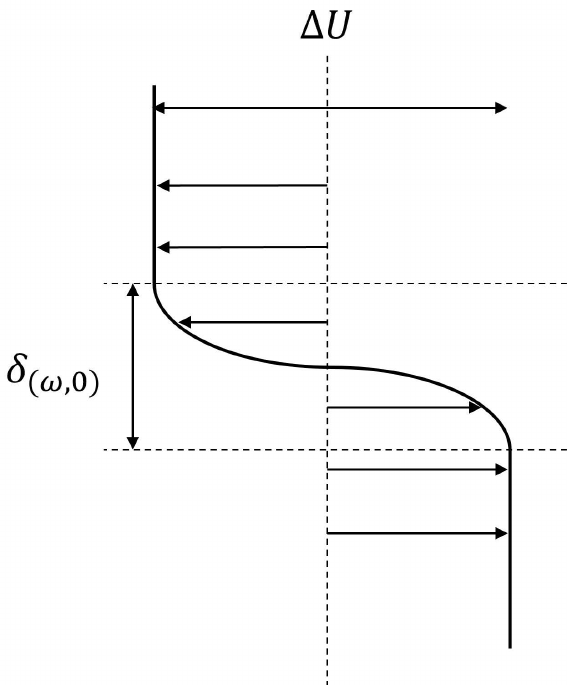}
\caption{Schematic of a temporally-evolving shear layer.}
\label{fig:temporal_shear_layer}
\end{figure}

The temporal shear layer data are from Ref. \cite{huang2021determining}.
Figure \ref{fig:temporal_shear_layer} is a schematic of the flow.
Two miscible fluids with velocities of equal magnitudes but opposite signs are brought together. 
The flow is homogeneous in the two horizontal directions and evolves in time. 
The initial velocity profile is given by
\begin{equation}
    U = - \frac{\Delta U}{2} \tanh \left( \frac{2(y-y_0)}{\delta_{(\omega,0)}} \right),
\end{equation}
where $\Delta U$ represents the difference in the velocity between the two fluids, $y_0=0$ is at the center of the domain, and $\delta_{(\omega,0)}$ is the initial vorticity thickness defined as $\delta_{\omega} \equiv \Delta U/(d \langle u \rangle/dy)_{max}$. 
The initial Reynolds number of the flow is $Re=\Delta U \delta_{(\omega,0)}/\nu=640$.
The momentum thickness is defined as $\delta_\theta \equiv \int_{-\infty}^{\infty}(1/4-\langle u \rangle^2) dy$.
The flow evolves temporally.
We reserve the flow at $t\Delta U/\delta_{(\omega,0)}=$ 5, 13, 15.5, 31, 49 for training and $t\Delta U/\delta_{(\omega,0)}=$ 46.5, 57 for testing.
Further details about the computational setup of the case can be found in Refs. \cite{jain2022second,huang2021determining} and are not repeated here for brevity.

\section{Results}
\label{sec:result}

We train against channel flow data and present the results in Sec. \ref{sub:chann}.
In Sec. \ref{sub:discussion}, we discuss the TBNN and PIML results.
In Sec. \ref{sub:generalization}, we apply the FIML model to other flows to test its generalizability.
Last, we train against both channel flow and mixing layer flow data to include another basic calibration.
The results are presented in section \ref{sub:C+M}.

Before we proceed, we would like to remark on the presentation of the results. 
For FIML, we present only the mean flow results, as the method does not concern any intermediate variables. 
For TBNN and PIML, we first present the target variable of machine learning and only examine the mean flow if ML yields good predictions for the target variables. 
It is not advisable to skip the machine learning variable(s): if there were errors in the mean flow, it would not be clear whether the culprit is the ML or velocity propagation.

\subsection{Channel flow}
\label{sub:chann}

The results are organized into 3 sections for TBNN, PIML, and FIML, respectively. 
Again, we train against data at $Re_\tau=180$, 590, 1000, and 2000.
Testing results will be shown at three Reynolds numbers: $Re_\tau=180$, one of the training conditions, $Re_\tau=395$, a condition that can be interpolated from the training set, and $Re_\tau=5200$, a condition that must be extrapolated.
There is an extended logarithmic layer at $Re_\tau=5200$, allowing us to test if the model has learned the LoW.

\subsubsection{TBNN}

The output $b_{ij}$ requires information on turbulent kinetic energy, which is not available in the SA model. As a result, TBNN, at least in its form as presented in Refs. \cite{ling2016reynolds,parish2016paradigm}, cannot be applied to the SA model.

\begin{figure}
\centering
\includegraphics[width=.55\textwidth]{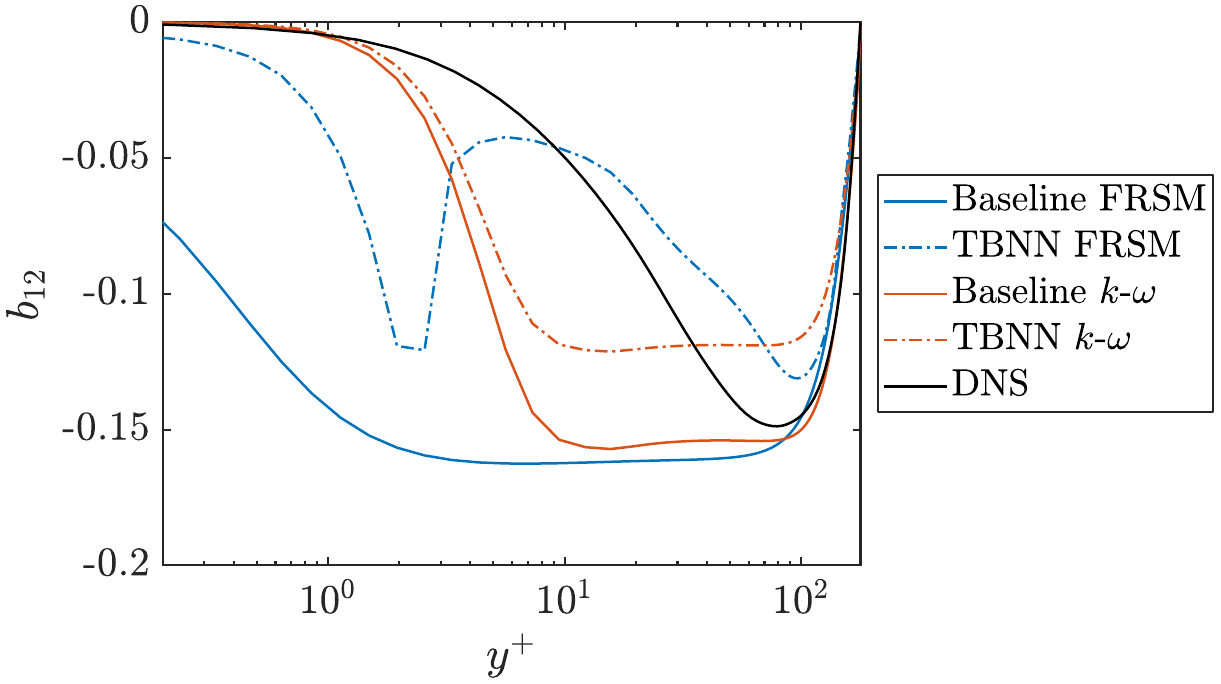}
\caption{Profiles of $b_{12}$ at $Re_\tau=180$. The black solid line is the DNS.
The colored solid lines are the results of the baseline RANS.
The dot-dashed lines are TBNN results.}
\label{fig:tbnn_180}
\end{figure}

\begin{figure}
\centering
\includegraphics[width=.75\textwidth]{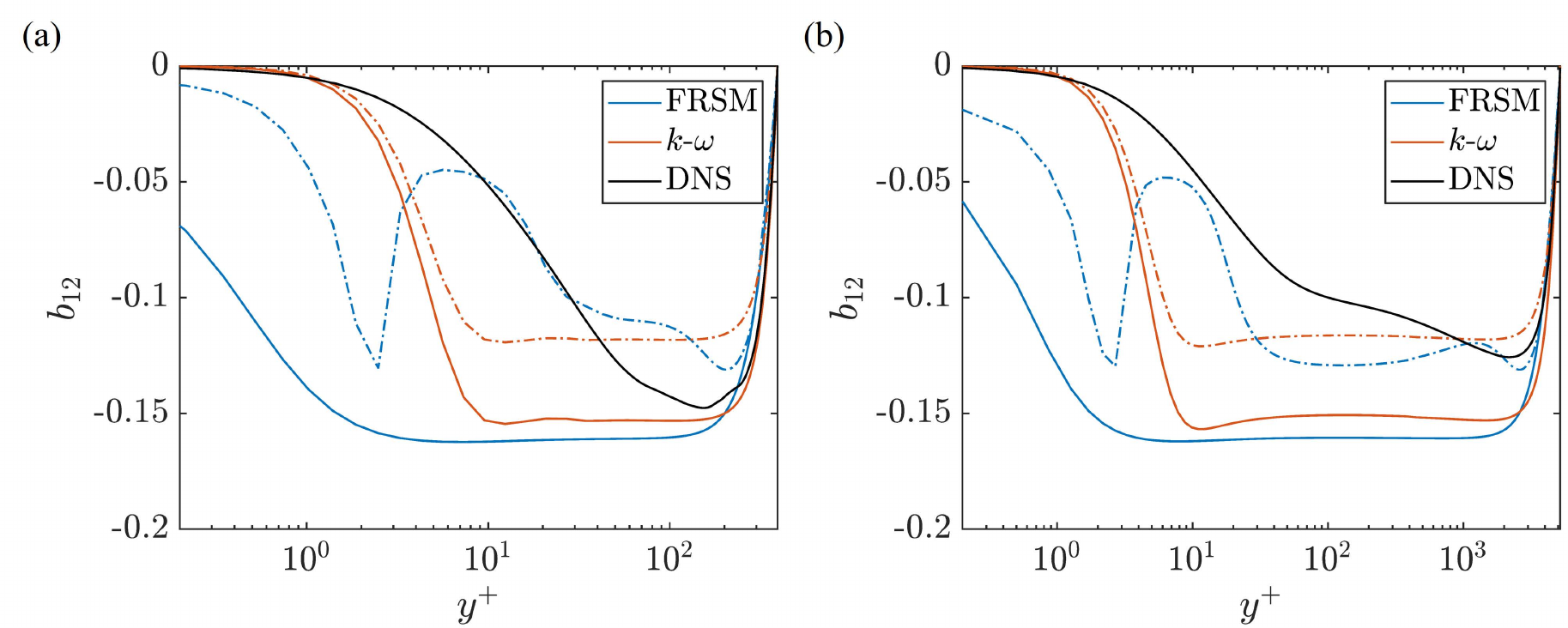}
\caption{Same as Fig. \ref{fig:tbnn_180} but at (a) $Re_\tau=395$, (b) $Re_\tau=5200$. Again, the colored solid lines are the results of the baseline RANS models;
the dot-dashed lines are TBNN results.}
\label{fig:tbnn_channel}
\end{figure}

Figure \ref{fig:tbnn_180} displays the TBNN-predicted $b_{12}$ at $Re_{\tau}=180$, with results from two baseline models included for comparison. Notably, the baseline SSG FRSM model and the baseline Wilcox $k$-$\omega$ model tend to over-predict $\left|b_{12}\right|$ in the wall layer. TBNN's augmentation to the Wilcox $k$-$\omega$ model maintains the overall behavior of $\left|b_{12}\right|$ but brings the profile closer to the DNS. The augmentation to the SSG FRSM alters the overall behavior of $b_{12}$ and results in a closer agreement between RANS and DNS at all $y$ locations. However, an unphysical peak emerges in the viscous sublayer, which will be discussed in Sec. \ref{sub:discussion}.
Figure \ref{fig:tbnn_channel} presents results at $Re_\tau=395$ and $5200$, showing similarities to those in Fig. \ref{fig:tbnn_180} at $Re_\tau=180$. The observed oscillations in the profiles at $Re_\tau=5200$ are expected, as noted in \cite{fang2020neural}.

\begin{figure}
\centering
\includegraphics[width=.98\textwidth]{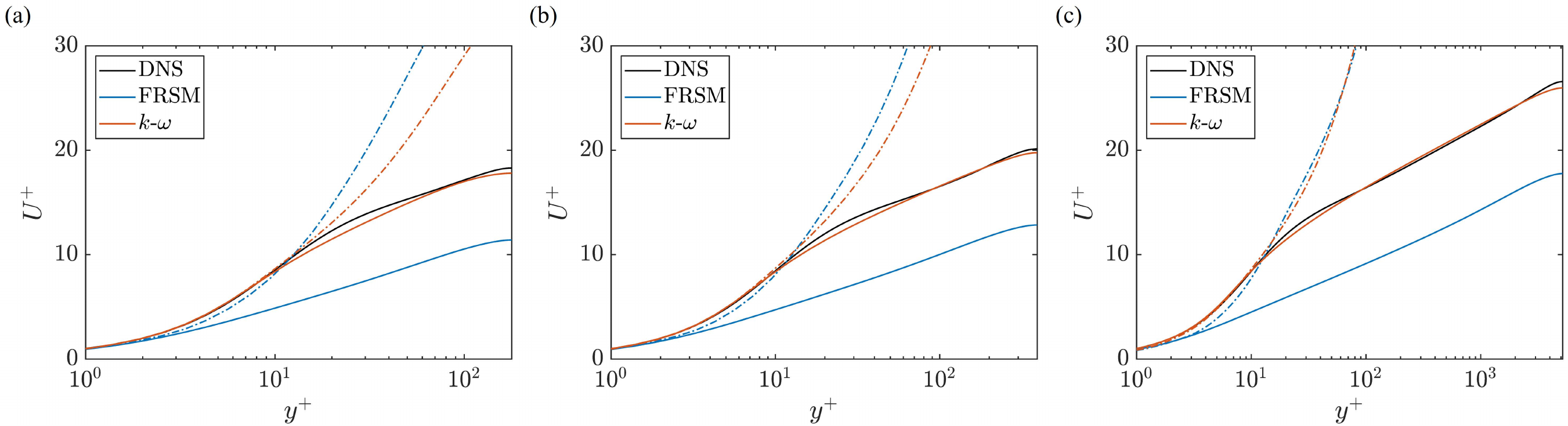}
\caption{Velocity profiles by integrating TBNN predicted $R_{12}$ through Eqs \ref{eq:inte} and \ref{eq:R12}. Colored solid lines are the results of the baseline RANS model. Colored dot-dashed lines are the results of TBNN. 
(a) $Re_\tau=180$, (b) $Re_\tau=395$, (c) $Re_\tau=5200$.}
\label{fig:TBNN_U}
\end{figure}

\begin{figure}
\centering
\includegraphics[width=.4\textwidth]{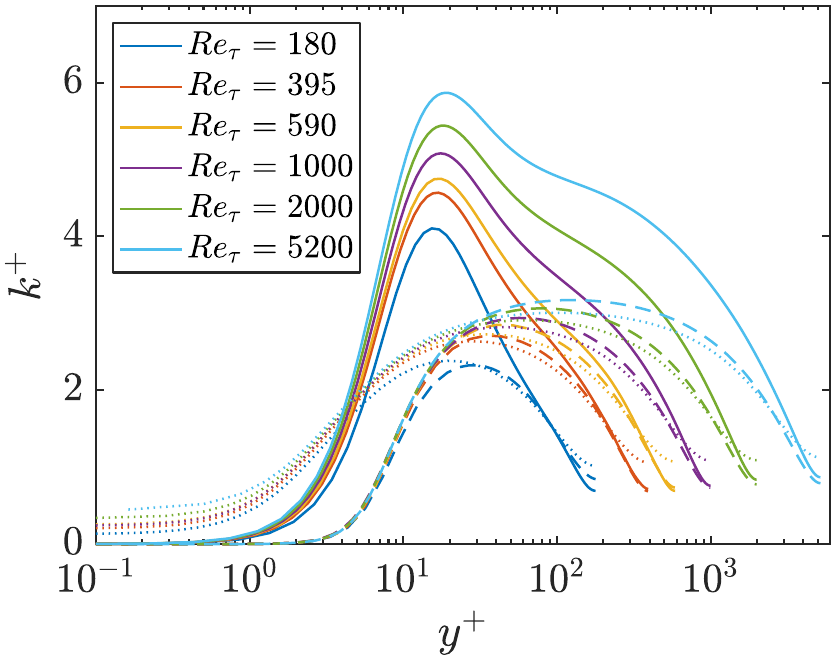}
\caption{Turbulent kinetic energy in DNS and the two baseline RANS models.
Here, different colors are used to distinguish different Reynolds numbers.
Solid lines are for DNS, dashed lines are for $k$-$\omega$, and dotted lines are for FRSM.}
\label{fig:tke_dns}
\end{figure}

Multiplying $b_{12}$ by $k$ and integrating the mean momentum equation yields the mean velocity profile:
\begin{equation}
U=\int_{0}^{y}\frac{dU}{dy} dy=\int_{0}^{y} \frac{1}{\nu}\left[(\tau_w/\rho)(1-y/\delta)+R_{12}\right]~dy,
\label{eq:inte}
\end{equation}
where
\begin{equation}
R_{12}=2b_{12}k_\text{Baseline RANS}.
\label{eq:R12}
\end{equation}
Figure \ref{fig:TBNN_U} illustrates the mean flow results. Although TBNN leads to improved $b_{12}$ predictions, these improvements do not necessarily translate into enhancements in the mean flow, primarily due to disparate values of $k$ in RANS and DNS, as depicted in Fig. \ref{fig:tke_dns}.

\subsubsection{PIML}

\begin{figure}
\centering
\includegraphics[width=.55\textwidth]{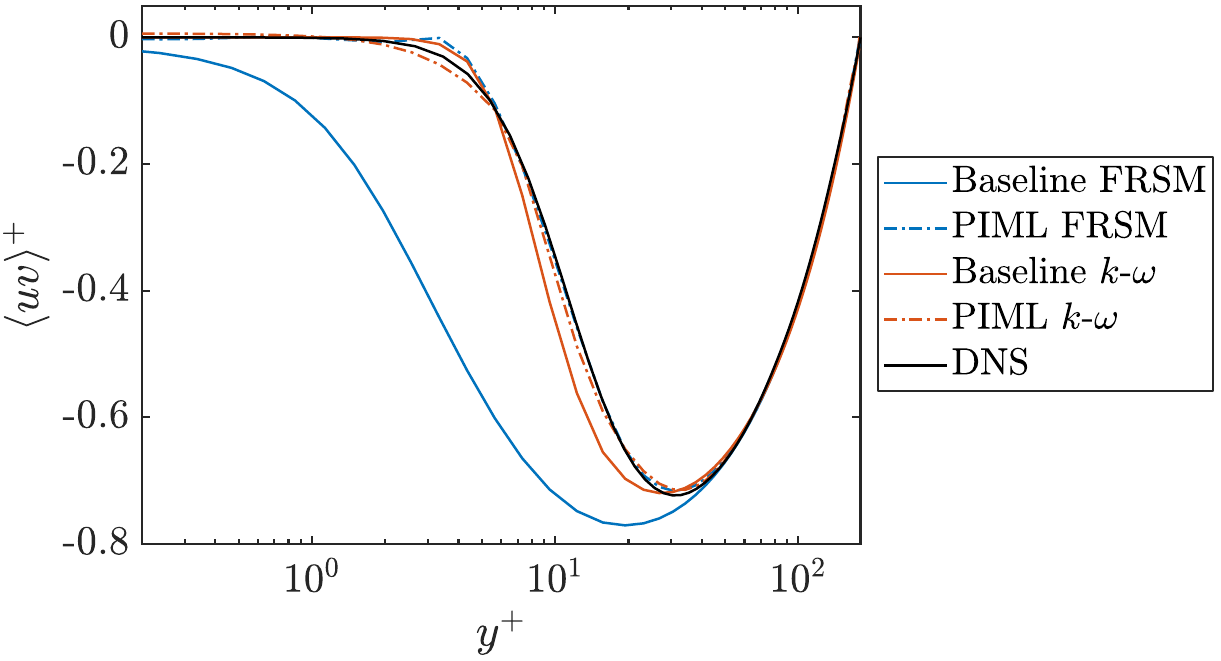}
\caption{The Reynolds shear stress $\langle uv \rangle$ at $Re_\tau=180$.
Colored solid lines are the results of the baseline RANS model. Colored dot-dashed lines are the results of PIML.}
\label{fig:PIML_180}
\end{figure}

Figure \ref{fig:PIML_180} illustrates the profiles of $R_{12}$ at $Re_\tau=180$. The baseline $k$-$\omega$ result is already very close to the DNS, while the SSG FRSM result overestimates the Reynolds shear stress. PIML's augmentation improves the $R_{12}$ prediction for both SSG FRSM and Wilcox $k$-$\omega$, resulting in nearly perfect $R_{12}$ profiles for both models.

\begin{figure}
\centering
\includegraphics[width=.75\textwidth]{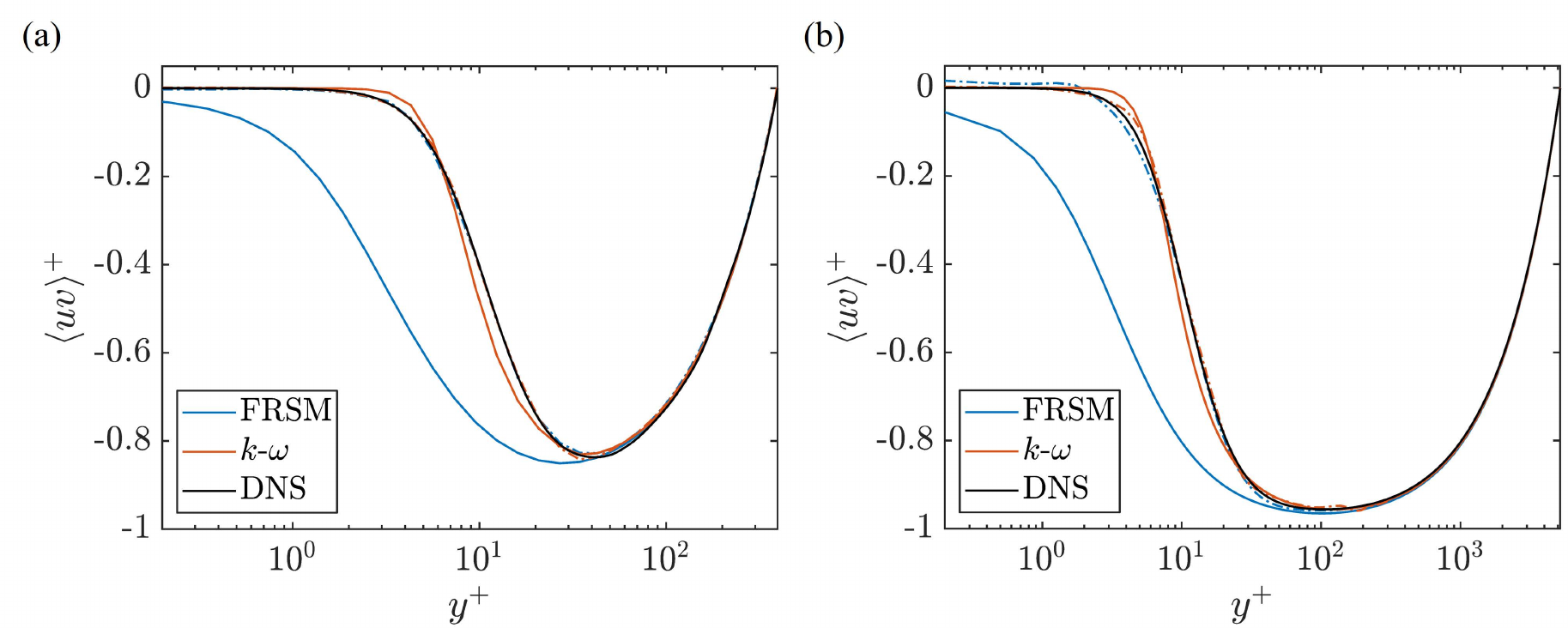}
\caption{Same as Fig. \ref{fig:PIML_180} but at (a) $Re_\tau=395$ and (b) $Re_\tau=5200$.}
\label{fig:PIML_channel}
\end{figure}

\begin{figure}
\centering
\includegraphics[width=.99\textwidth]{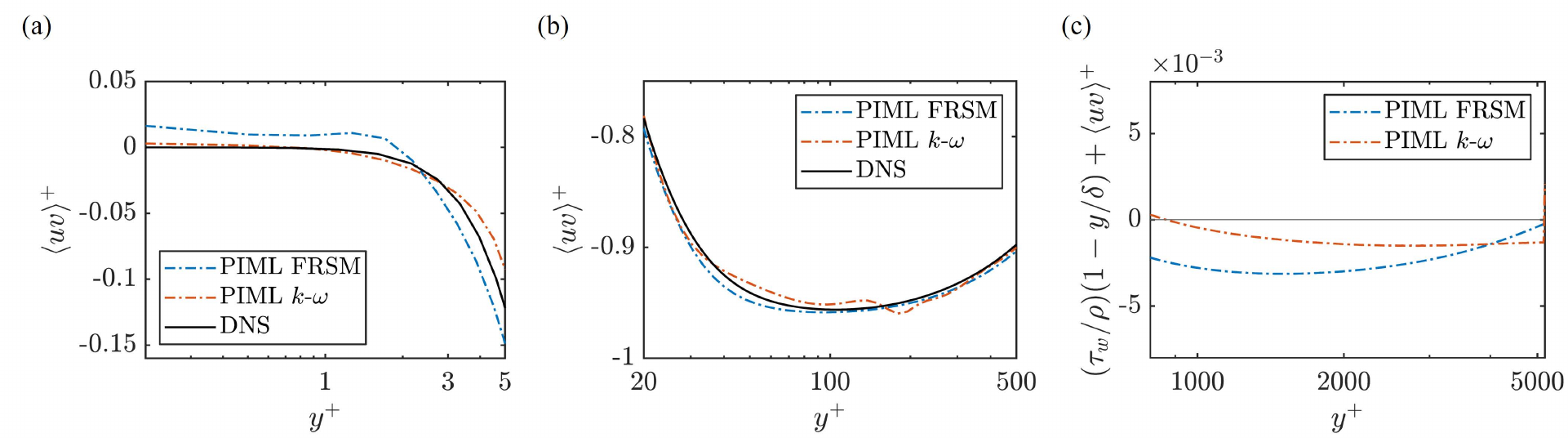}
\caption{A zoom-in view of PIML's predictions at $Re_{\tau}=5200$. (a) for $0.2\leq y^+\leq5$, (b) for $10\leq y^+ \leq500$, (c)$\left[(\tau_w/\rho)(1-y/\delta)+\langle uv \rangle^+\right]$, i.e., $dU^+/dy^+$ for $800\leq y^+\leq5200$ .}
\label{fig:PIML_channel_zoom}
\end{figure}

Figure \ref{fig:PIML_channel} displays the results at $Re_\tau=395$ and 5200. Although it appears that PIML's augmentation enhances the predictions of $R_{12}$ at both Reynolds numbers, closer inspection in Fig. \ref{fig:PIML_channel_zoom} reveals some inaccuracies. Specifically, PIML's prediction of $R_{12}$ is slightly positive in the viscous sublayer at $Re_\tau=5200$, accompanied by unphysical undulations around $y\approx 100$. Furthermore, $dU^+/dy^+=1-y/\delta+\left<u'v'\right>^+$ is slightly negative in the wake layer. These unphysical behaviors, as we will discuss shortly, significantly impact the mean flow.

\begin{figure}[htb!]
\centering
\includegraphics[width=.99\textwidth]{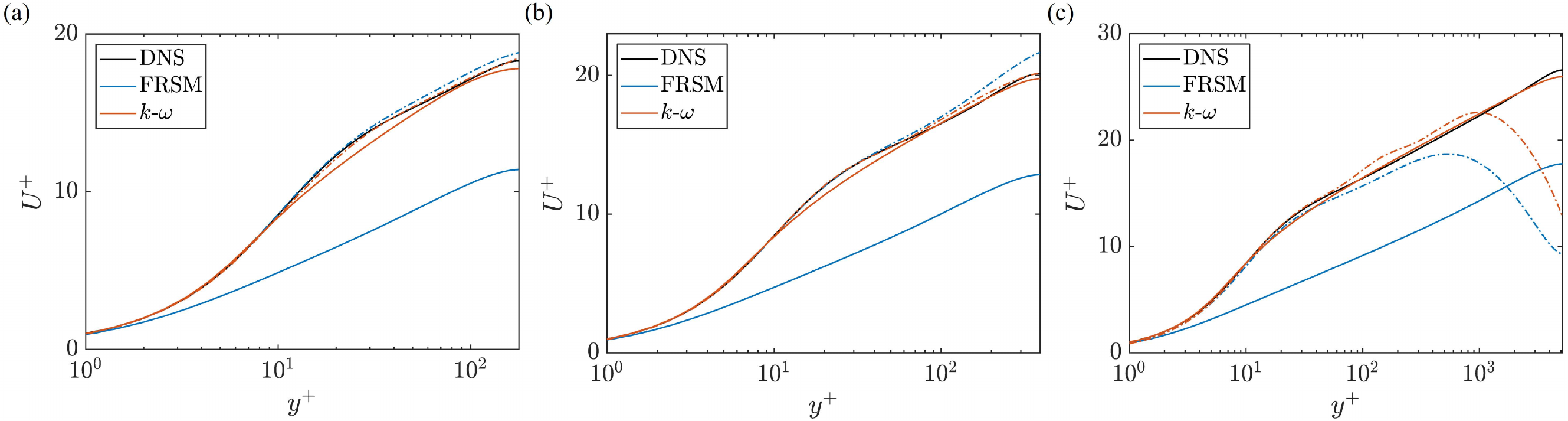}
\caption{Velocity profiles by integrating PIML predicted $R_{12}$ according to Eq. \ref{eq:inte}. Colored solid lines are the results of the baseline RANS models. Colored dot-dashed lines are the results of PIML. 
(a) $Re_\tau=180$, (b) $Re_\tau=395$, (c) $Re_\tau=5200$.}
\label{fig:PIML_U}
\end{figure}

Integrating the mean momentum equation as per Eq. \eqref{eq:inte} gives the mean flow. Figure \ref{fig:PIML_U} presents the results. In comparison to the TBNN results in Fig. \ref{fig:TBNN_U}, the PIML results exhibit more favorable outcomes here. PIML's augmentation improves the mean flow predictions of the two baseline models at both $Re_\tau=180$ and 395. For SSG FRSM, PIML's augmentation acts as a wall treatment. However, PIML's augmentation proves detrimental when extrapolating, as depicted in Fig. \ref{fig:PIML_U} (c), due to the errors shown in Fig. \ref{fig:PIML_channel_zoom}. This is consistent with the findings in Ref. \cite{guo2021computing}, where it was demonstrated that velocity propagation incurs significant errors.

\subsubsection{FIML}

\begin{figure}
\centering
\includegraphics[width=.4\textwidth]{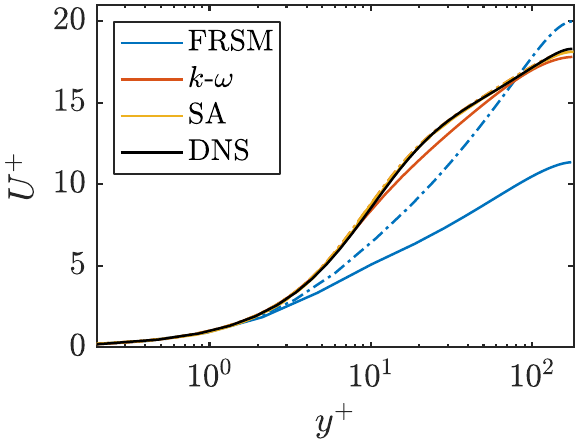}
\caption{Velocity profiles at $Re_\tau=180$. Here, colored solid lines represent baseline results and colored dot-dashed lines represent FIML results.}
\label{fig:FIML_180}
\end{figure}

\begin{figure}[htb!]
\centering
\includegraphics[width=.75\textwidth]{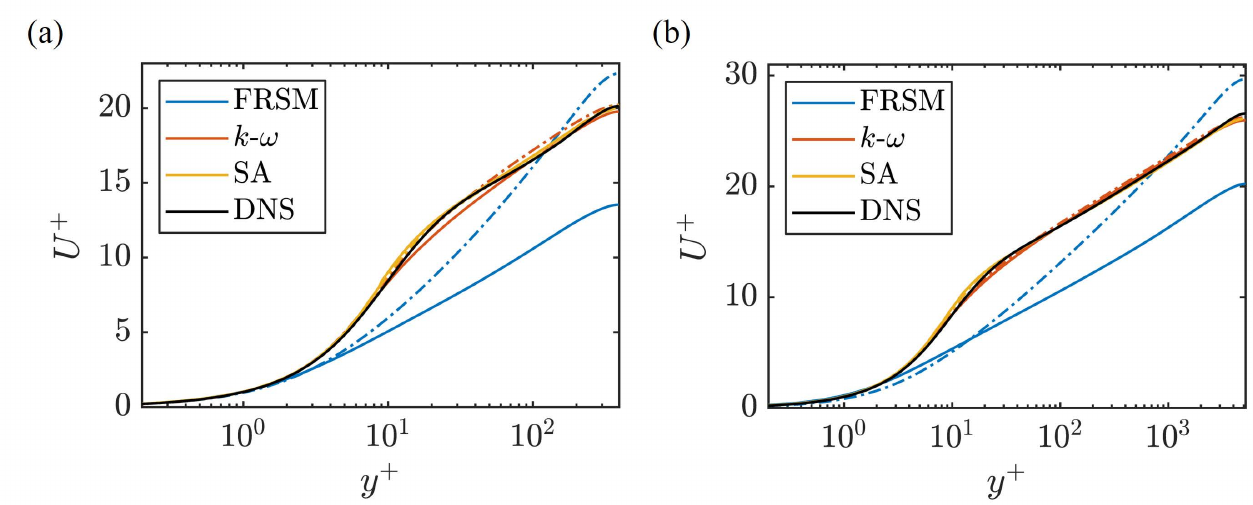}
\caption{Same as Fig. \ref{fig:FIML_180} but at (a) $Re_\tau=395$, (b) $Re_\tau=5200$.}
\label{fig:FIML_channel}
\end{figure}

FIML is applicable to all baseline models, and the training specifically addresses the mean flow itself. Figure \ref{fig:FIML_180} presents the mean flow results at $Re_{\tau}=180$. The baseline SA and Wilcox $k$-$\omega$ models already provide good mean flow predictions in plane channels. For these two models, FIML's augmentations are largely neutral, with slight improvements observed in the buffer layer. 
The baseline SSG FRSM lacks wall damping.
FIML's augmentation leads to a different log law slope.
While FIML's augmentation brings the RANS profile closer to the DNS profile, the solution does not achieve the correct layered structure.
Figure \ref{fig:FIML_channel} depicts the results at $Re_\tau=395$ and 5200, showing similarities to those in Fig. \ref{fig:FIML_180}. For SA and Wilcox $k$-$\omega$, where there is already a wall treatment, FIML's augmentation brings the profile closer to the DNS in the buffer layer. In the case of SSG FRSM, lacking a wall treatment, FIML changes the log layer slope, bringing the RANS profiles closer to the DNS.
We see that FIML's augmentations are non-detrimental.
This is different from the results in \cite{rumsey2022search}, a discussion of which is postponed to section \ref{sub:discussion}.

\vspace{5mm}

Table \ref{tab:result} summarizes the results in Sec. \ref{sub:chann}.
The data-based augmentations due to TBNN prove to be detrimental both inside and outside the training set.
The augmentations due to PIML are beneficial inside the training set but are detrimental outside it.
The augmentations due to FIML are non-detrimental inside or outside the training set.
However, we see that when the baseline model lacks the right physics, FIML does not necessarily learn that physics.

\begin{table}
\caption{A summary of channel flow results}
\label{tab:result}
\centering
\begin{tabular}{l|lll}
\hline
     & \multicolumn{1}{l|}{SA}  & \multicolumn{1}{l|}{Wilcox $k$-$\omega$}             & SSG FRSM                                          \\ \hline
TBNN & \multicolumn{1}{l|}{N/A} & \multicolumn{2}{l}{Improved $b_{12}$, detrimental to $U$ inside and outside the training set}                \\ \hline
PIML & \multicolumn{1}{l|}{N/A} & \multicolumn{2}{l}{Improved $R_{12}$, beneficial and detrimental to $U$ inside and outside the training set} \\ \hline
FIML & \multicolumn{2}{l|}{Neutral in log layer, beneficial in the buffer layer}  & Beneficial, but did not capture the right physics  \\ \hline
\end{tabular}
\end{table}

\subsection{Further discussion of TBNN and PIML}
\label{sub:discussion}

\begin{figure}
\centering
\includegraphics[width=.75\textwidth]{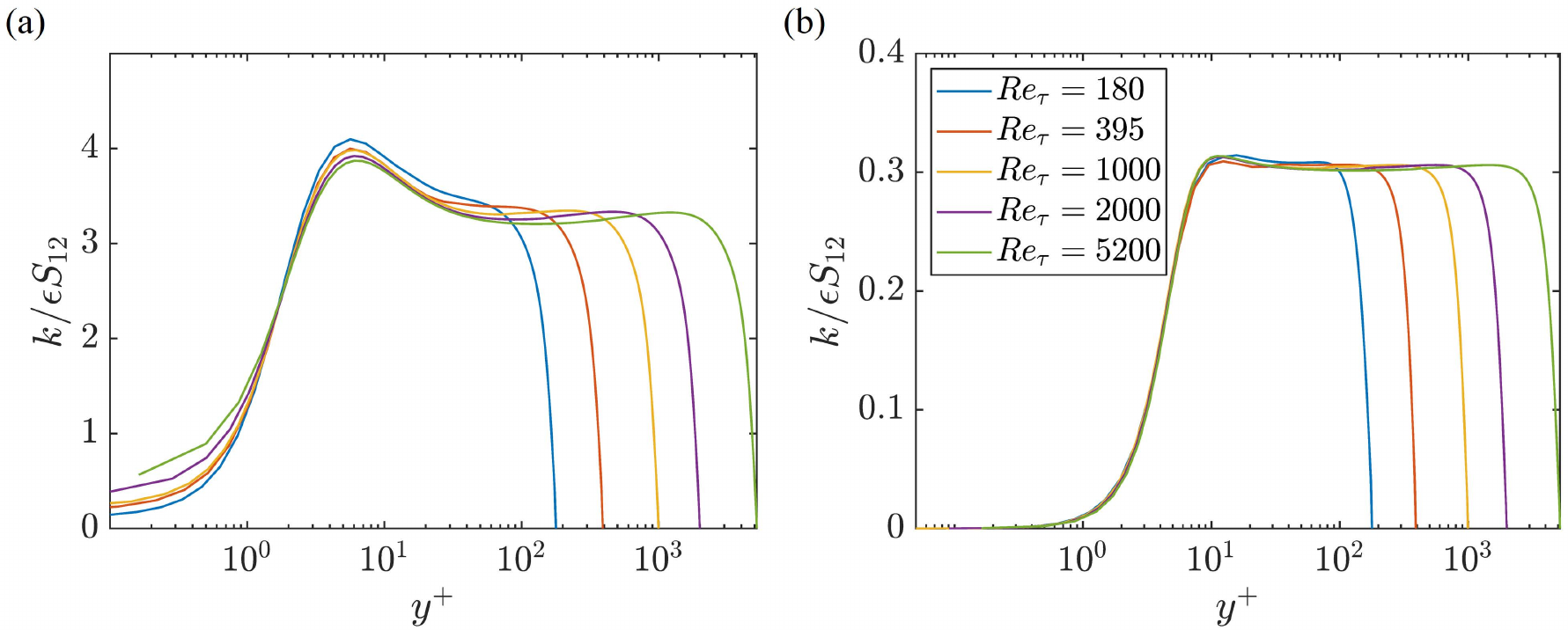}
\caption{$k/\epsilon \partial{U}/\partial{y}$ due to (a) the baseline FRSM and (b) the baseline Wilcox $k$-$\omega$ model.}
\label{fig:tbnn_input}
\end{figure}

\begin{figure}
\centering
\includegraphics[width=0.77\textwidth]{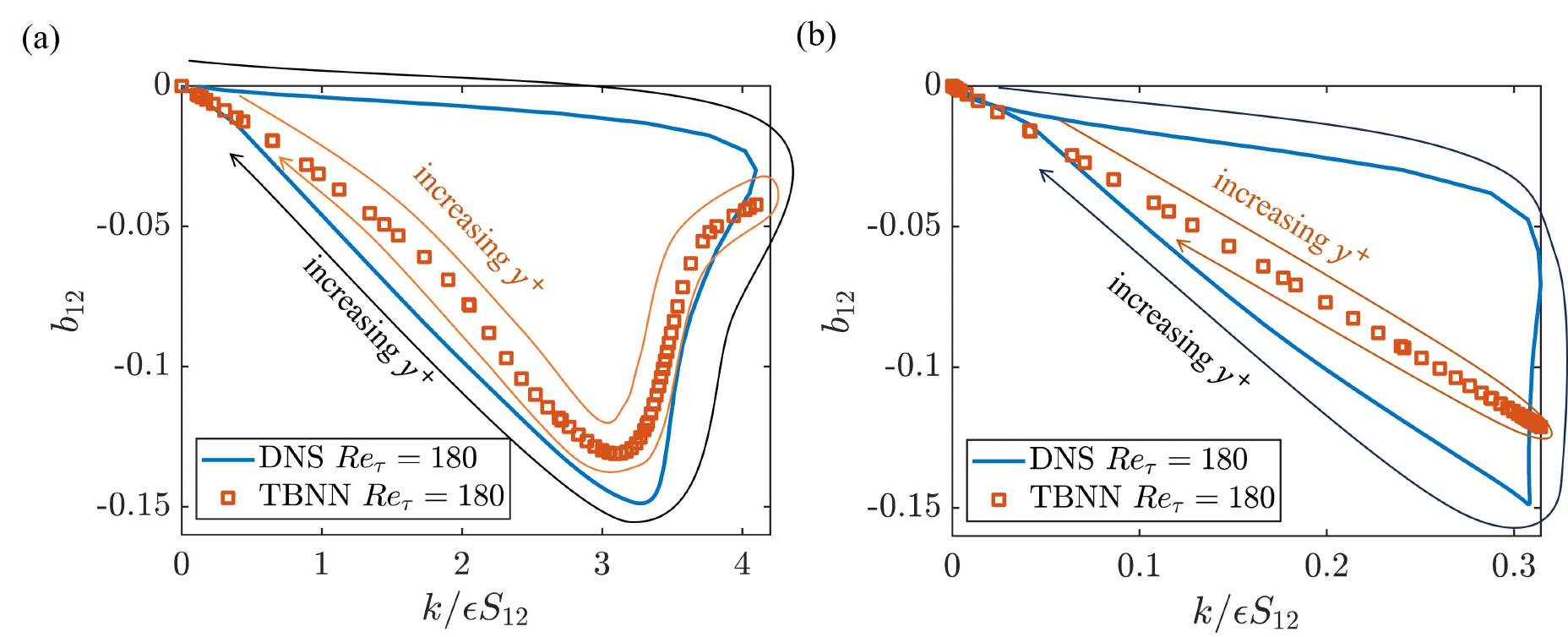}
\caption{Desired $b_{12}$ and TBNN predicted $b_{12}$ as a function of $k/\epsilon \partial{U}/\partial{y}$ for (a) FRSM and (b) $k$-$\omega$.
We indicate the direction of increasing $y^+$.}
\label{fig:tbnn_input_2}
\end{figure}

For TBNN, the network's inputs consist of invariants derived from the strain rate and rotation tensor. In the context of plane channels, the strain rate tensor and the rotation tensor have only one non-zero component, which is $dU/dy$. Therefore, any non-zero input to the network is a result of a non-zero $dU/dy$. In Fig. \ref{fig:tbnn_input}, we illustrate $k/\epsilon \partial{U}/\partial{y}$ in the baseline FRSM and Wilcox $k$-$\omega$ model. It is evident that $k/\epsilon \partial{U}/\partial{y}$ increases with $y^+$ in the inner layer and decreases with $y^+$ in the outer layer. Consequently, a given $k/\epsilon \partial{U}/\partial{y}$ value appears twice: once in the inner layer and a second time in the outer layer. A direct consequence is that the target variable of TBNN, $b_{12}$, is not a single-valued function of any inputs derived from $k/\epsilon \partial{U}/\partial{y}$.
This is clear from Fig. \ref{fig:tbnn_input_2}, where we display $b_{12}$ as a function of $k/\epsilon \partial{U}/\partial{y}$ at $Re_\tau=180$. 
The results at other Reynolds numbers are similar and therefore are not shown here for brevity.
However, a neural network, particularly in the form presented in Ref. \cite{ling2016reynolds}, is inherently single-valued. When trained to predict $b_{12}$ as a function of variables derived from $k/\epsilon S_{12}$, the neural network tends to compromise between multiple $b_{12}$ values. The predictions of the trained networks, as a function of $k/\epsilon dU/dy$, are demonstrated in Fig. \ref{fig:tbnn_input_2}. For FRSM, the predicted $b_{12}$ exhibits non-monotonic behavior with respect to $k/\epsilon dU/dy$. This non-monotonic behavior is responsible for the two peaks observed in $b_{12}$ profiles in Figs. \ref{fig:tbnn_180} and \ref{fig:tbnn_channel}. 
The above explains why TBNN is detrimental.

\begin{figure}[htb!]
\centering
\includegraphics[width=.75\textwidth]{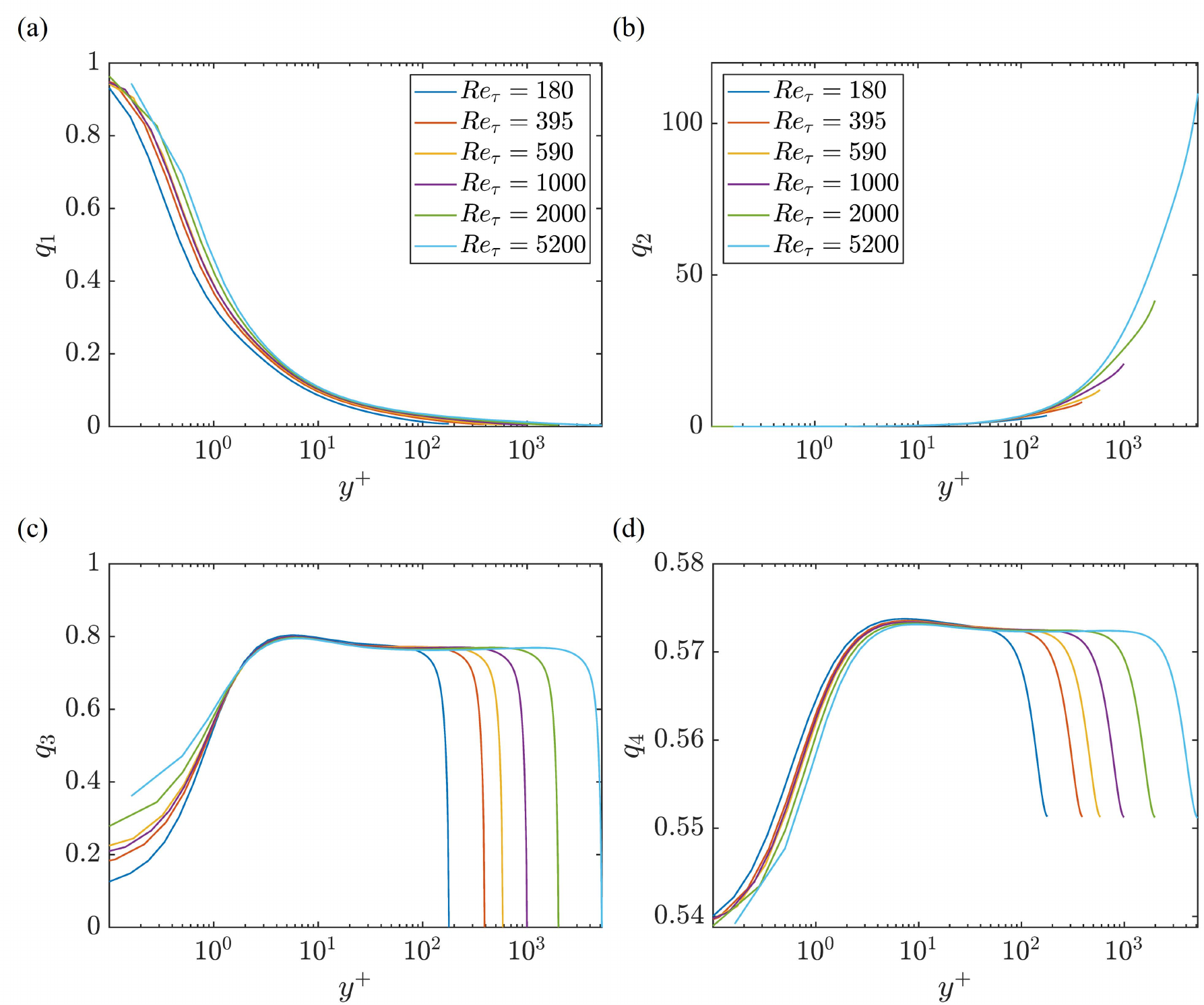}
\caption{Normalized inputs of PIML calculated from FRSM (a) $k$, (b) $\frac{\sqrt{k}d}{50\nu}$, (c) $k/\epsilon$, (d) $||\overline{u'_iu'_j}||$. The normalization process follows that in Ref. \cite{wang2017physics} and given in Table \ref{table:piml}.}
\label{fig:piml_input}
\end{figure}

Next, we consider PIML.
In the case of plane channel flow, four of the O(10) inputs to the network are non-zero. Figure \ref{fig:piml_input} visually represents these four non-zero inputs. Notably, inputs $q_1$ and $q_2$ exhibit monotonic behavior as functions of $y^+$. This characteristic allows a neural network to accurately approximate the target variable as a function of either $q_1$, $q_2$, or a combination of both. Consequently, the network is capable of making improvements within the training dataset.
However, since the Reynolds stress must be integrated to obtain the mean flow, PIML is susceptible to errors stemming from the integration process. As noted by Wu et al., depending on the chosen integration method, there could be a substantial 35\% error in the mean flow \cite{wu2019Reynolds}.
The above explains why PIML is beneficial inside the training dataset but detrimental outside.

\subsection{Generalization of FIML}
\label{sub:generalization}

We implement FIML's augmentation of the $k-\omega$ model in OpenFOAM \cite{weller1998tensorial} and assess its performance in scenarios involving a mixing layer and a backward-facing step. It is important to note that we do not retrain for these two cases. The objective is to scrutinize whether FIML's augmentation proves detrimental outside the training dataset.
The test in this subsection is limited and more comprehensive tests fall outside the scope of this work.
Our methodology follows the steps outlined in Ref. \cite{rumsey2022search}. We first allow the baseline model to converge, obtaining the initial guess for the $\beta$ field from the converged flow field. Subsequently, we let the augmented model converge. The detailed setups for the two cases can be found in Ref. \cite{NASATMR} and are not reiterated here for brevity.

\begin{figure}
\centering
\includegraphics[width=.95\textwidth]{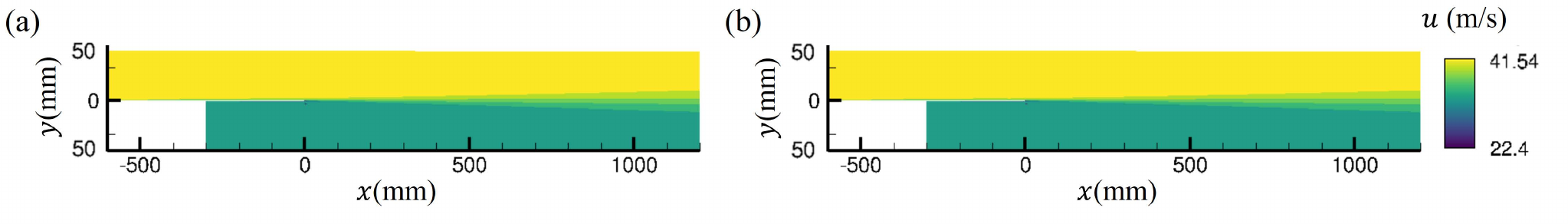}
\caption{Velocity contours of the mixing-layer flow. (a) Baseline Wilcox $k$-$\omega$ result, (b) FIML-augmented model result.}
\label{fig:mixing_layer_contour}
\end{figure}

\begin{figure}
\centering
\includegraphics[width=.99\textwidth]{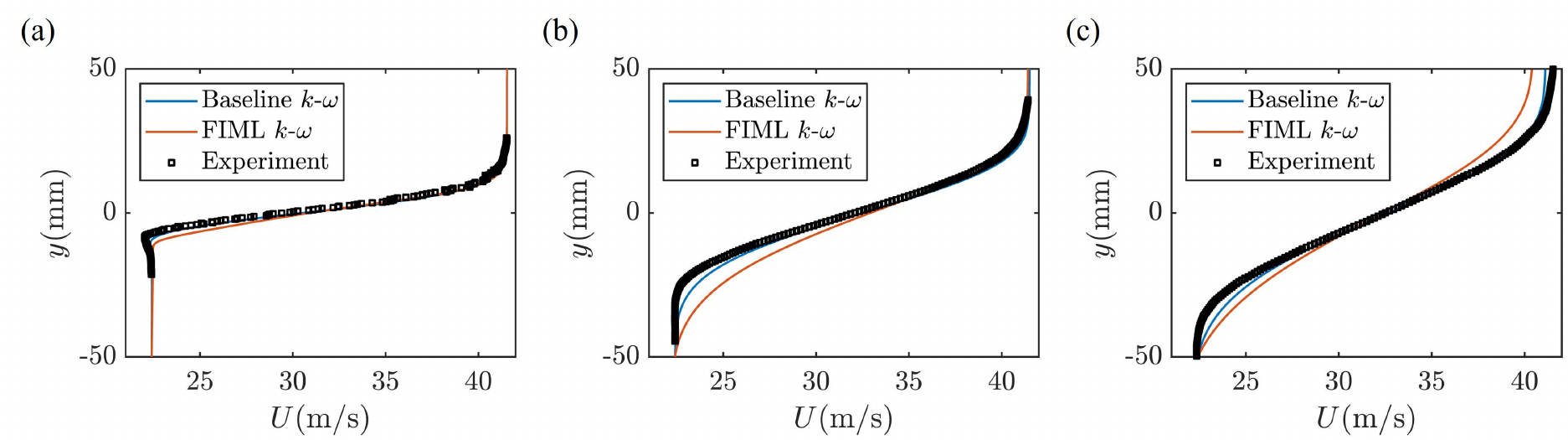}
\caption{Velocity profiles in the shear layer at $x=$ (a) 200mm, (b) 650mm, and (c) 950mm.}
\label{fig:spacial_ml_line}
\end{figure}

\begin{figure}[htb!]
\centering
\includegraphics[width=.75\textwidth]{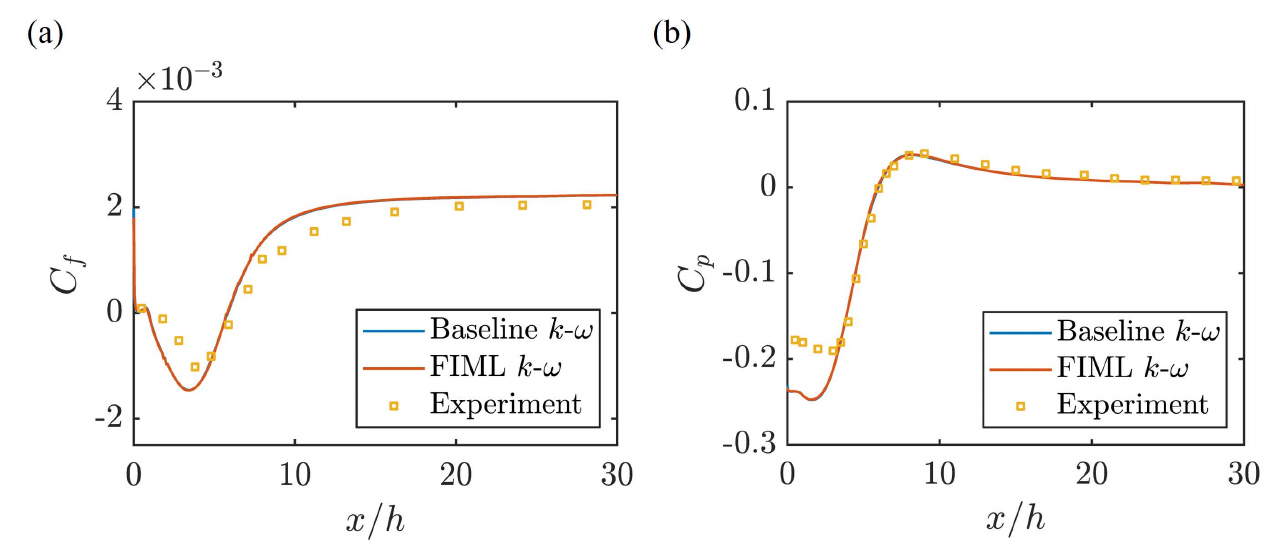}
\caption{(a) Skin friction coefficient  $C_f$ and (b) pressure coefficient $C_p$ of the backward-facing-step.}
\label{fig:bfs_line}
\end{figure}

The results for the spatial mixing layer are depicted in Figs. \ref{fig:mixing_layer_contour} and \ref{fig:spacial_ml_line}. Figure \ref{fig:mixing_layer_contour} displays the velocity contour of both the baseline model and the FIML-augmented model. Meanwhile, Fig. \ref{fig:spacial_ml_line} illustrates the velocity profiles at various downstream locations, comparing the experimental data from Ref. \cite{delville1989analysis}, the baseline Wilcox $k$-$\omega$ model, and the FIML-augmented model.
The contour plot in Fig. \ref{fig:mixing_layer_contour} shows no noticeable differences between the baseline and FIML. 
Differences between the two are more clear from the line plots in Fig. \ref{fig:spacial_ml_line}.
The results of the baseline model are already very accurate.
FIML's augmentation gives rise to enhanced mixing, which negatively impacts the results.
The effect is, however, not as concerning as in Ref. \cite{rumsey2022search}.
Figure \ref{fig:bfs_line} shows the results for the backward-facing step case.
We plot the skin friction coefficient $C_f$ and the pressure coefficient $C_p$ for the baseline model and the FIML-augmented model.
We see that FIML's augmentation does not affect the result. 
Most importantly, it is non-detrimental.

\subsection{Incorporating other basic calibrations}
\label{sub:C+M}

In addition to the LoW, other basic calibrations of RANS models include decaying isotropic turbulence, shear layers, etc.
In this section, we build on Sec. \ref{sub:chann} and train against both channel flow and temporally-evolving shear layers.

The shear flow is unsteady, applying FIML requires adjusting the RANS solution at many time steps.
This process can lead to ill-posedness \cite{srivastava2022generalizable}, and therefore FIML is not pursued here.
Furthermore, since TBNN is detrimental when trained against channel flow data, we do not pursue TBNN here either.
In this section, we will focus on PIML and the baseline $k$-$\omega$ model.

\begin{figure}
\centering
\includegraphics[width=.75\textwidth]{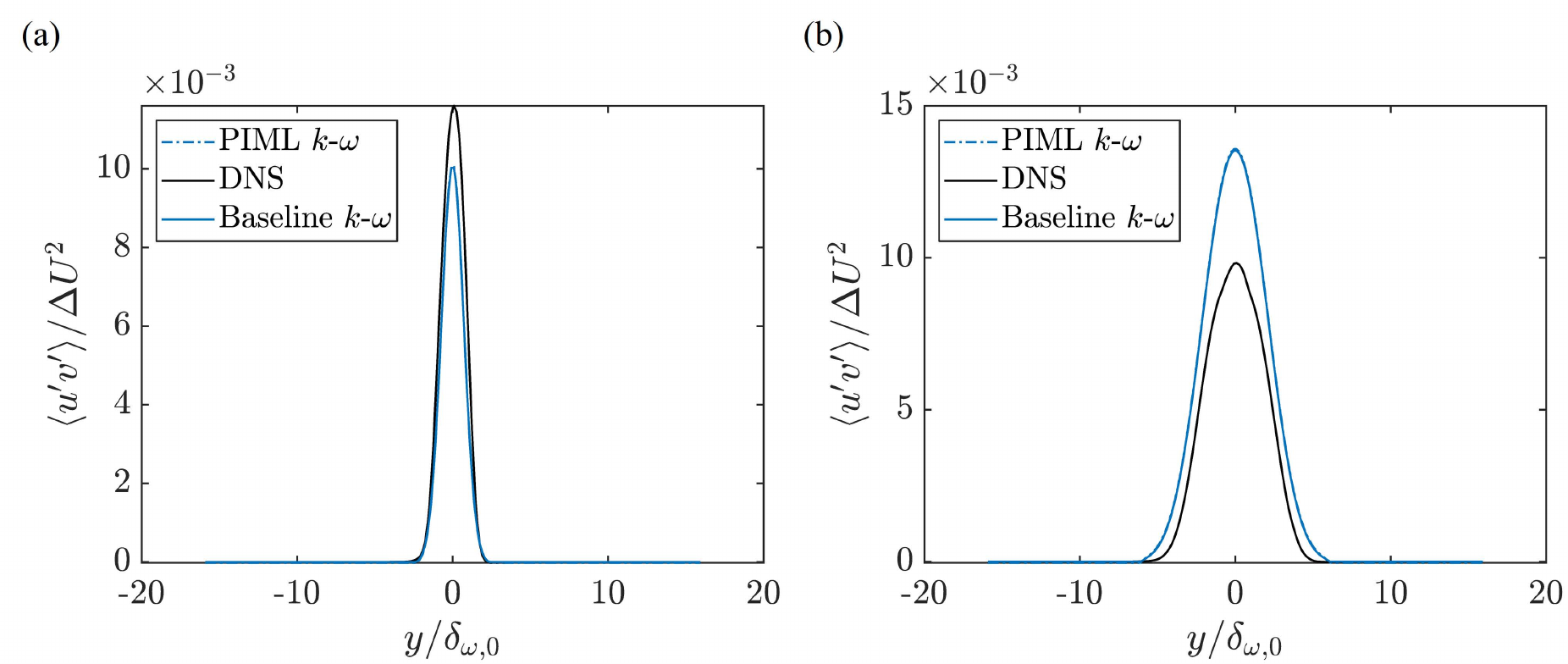}
\caption{Reynolds stress profile of mixing layer at (a) $t=15.5$, (b) $t=57$. The black solid line represents the DNS in Ref. \cite{huang2021determining}, the colored solid line is the result of the baseline $k$-$\omega$ model, and the colored dot-dashed line is the prediction from PIML.}
\label{fig:mixing_layer_piml}
\end{figure}

\begin{figure}
\centering
\includegraphics[width=.35\textwidth]{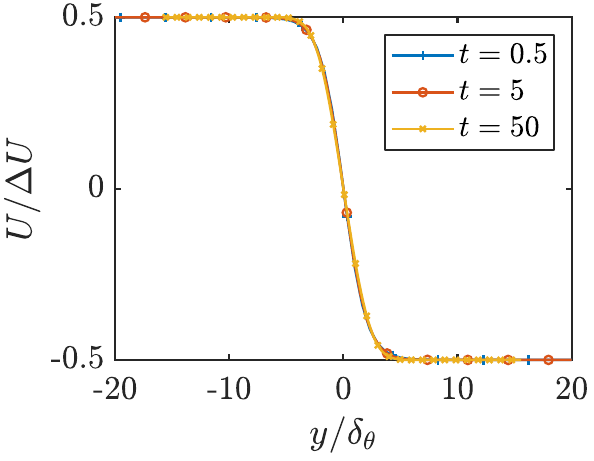}
\caption{Self-similarity of the temporal mixing layer: (a) velocity profile at various time steps.}
\label{fig:mixing_layer_self_similar}
\end{figure}

The results for the plane channel are similar to those presented in Sec. \ref{sec:result} and are not shown here for brevity.
We plot the Reynolds shear stress $\langle uv \rangle$ due to the baseline model, the DNS, and PIML at $t=15.5$ and 57 in Fig. \ref{fig:mixing_layer_piml}.
Here, time is normalized by $\Delta U$ and $\delta_{(\omega,0)}$.
We see that the baseline model underestimates and overestimates $\langle uv \rangle$ at the early and late stages of the flow and that PIML yields no improvement.
In the following, we explain why there is no improvement.
First, we note that the flow is roughly self-similar, i.e., $U/\Delta U$ is a function of $y/\delta_\omega$ only and not a function of time.
This is clear from Fig. \ref{fig:mixing_layer_self_similar}.
A direct consequence of self-similarity is that the locally normalized PIML inputs, i.e., $q_1$, $q_2$, $q_3$, and $q_4$, do not depend on time.
Next, we consider the error in the baseline model.
The error distribution in the baseline model necessitates PIML to predict a positive and a negative correction to the baseline model.
This is not possible with inputs that do not vary with time: improvements in the early stage will necessarily result in degradation in the late stage and vice versa.
As the error of the baseline $k$-$\omega$ already gives a rather balanced error distribution, PIML did not yield further improvements.

\section{Concluding remarks}
\label{sec:conclusion}
We apply TBNN, PIML, and FIML to the one-equation SA model, the two-equation $k$-$\omega$ model, and the seven-equation FRSM. TBNN and PIML as presented in Refs. \cite{ling2016reynolds,wang2017physics} directly modifies the Reynolds stress, and FIML as presented in Ref. \cite{parish2016paradigm,holland2019towards} modifies terms in the auxiliary transport equations. The SA and the $k$-$\omega$ models are eddy viscosity models, whereas FRSM is a Reynolds stress model. We train against plane channel flows and shear layer flows.
The SA and the $k$-$\omega$ model already yield accurate mean flow estimates for plane channel. The SSG FRSM, on the other hand, does not capture the mean flow in the buffer layer.

TBNN and PIML, at least in the forms as presented in Refs. \cite{ling2016reynolds,wang2017physics}, require TKE and dissipation for normalization purposes. As such information is not available in the SA model, TBNN and PIML are not applicable to the SA model. We first train against plane channel flows at various Reynolds numbers. Applying TBNN to the $k$-$\omega$ model and the SSG FRSM, we see unphysical but nevertheless improved predictions of $b_{12}$. However, the improvements in $b_{12}$ do not translate to the mean flow, and TBNN is detrimental both inside and outside the training dataset. (This conclusion applies only to the flows considered here.) Further analysis shows that TBNN yields degrading results here because the target variable is not a single-valued function of the inputs in this flow. Next, PIML yields improved $R_{12}$ predictions. However, due to the errors associated with velocity propagation, PIML is beneficial only at the Reynolds numbers inside the training dataset and is detrimental outside. These conclusions are independent of the baseline model. 

Training against plane channel flow data at various Reynolds numbers, FIML’s augmentations are largely neutral in the SA and the $k$-$\omega$ model with slight improvements in the buffer layer. Its augmentation to the SSG FRSM leads to a closer agreement between the RANS result and the DNS data, but such improvement is achieved via varying the von K{\'a}rm{\'a}n constant rather than a layer-structure of the mean flow (i.e., sublayer, buffer layer, log layer). The generalizability of FIML’s augmentation in the $k$-$\omega$ model is tested in backward-facing step and mixing-layer scenarios. The augmentation has essentially no impact on these two flows: we see slightly deteriorated results for the mixing layer and no noticeable change in the results for the backward-facing step. 

The scope of the present study is limited to three ML frameworks, three baseline RANS models, and two fluid flows. There have been further developments since the early publications in Refs. \cite{ling2016reynolds,wang2017physics,parish2016paradigm} of the three ML frameworks. Furthermore, there are other ML frameworks \cite{weatheritt2016novel,zhao2020rans,fang2023toward,bin2022progressive,rincon2023progressive}, other baseline RANS models, and other fluid flows. To enable the use of machine learning models in fluids engineering, many more comparative studies and validation \& verification studies are yet needed.

\section*{Acknowledgments}
Li and Yang acknowledge ONR contract number N000142012315 and AFOSR award number FA9550-23-1-0272.
Bin acknowledges NNSFC grant number 91752202.

\bibliography{sample}

\end{document}